\documentclass{aa}  

\usepackage{graphicx}
\usepackage{float}
\usepackage{amsmath,amssymb}
\usepackage{xcolor}
\usepackage{multirow} 

\usepackage{array} 
\usepackage{booktabs} 
\usepackage{colortbl}
\usepackage{tabularray}
\UseTblrLibrary{booktabs}
\SetTblrStyle{note}{font=\footnotesize, verb}

\usepackage{txfonts}
\begin{document}

   \title{Expansion kinematics of young clusters. II.\\NGC~2264 N \& S and Collinder 95 with HectoSpec}


   \author{Ishani Cheshire
          \inst{1,2}
          \and
          Joseph J. Armstrong\inst{1}\fnmsep\thanks{Corresponding author; jarmstrongastro@gmail.com}
          \and
          Jonathan C. Tan\inst{1,3}
          }

   \institute{Department of Space, Earth \& Environment, Chalmers University of Technology, SE-412 96 Gothenburg, Sweden
         \and
             Department of Astronomy, University of California, Berkeley, CA 94720-3411, USA
        \and
            Department of Astronomy, University of Virginia, Charlottesville, VA 22904, USA
             }


 
  \abstract
   {}
   {Studying the dynamical evolution of young clusters is crucial for a more general understanding of the star formation process. }
   {We took spectra of $>$600 candidate pre-main sequence (PMS) stars in several nearby young clusters (NGC 2264 N \& S, Collinder 95, and Collinder 359) using MMT/Hectospec. These spectra were analyzed for H$\alpha$ emission and lithium absorption, features indicative of low-mass young stellar objects (YSOs) still in their PMS evolution. 
   We complemented these samples with YSOs identified via \textit{Gaia} DR3 variability. In conjunction with \textit{Gaia} astrometry, these data enable an analysis of cluster structure, kinematics and ages. In particular, we searched for halos of YSOs around our targets to test models of young cluster dynamical evolution.}
   {For the NGC 2264 N \& S cluster pair we identified 354 YSOs, 
   while for Collinder 95 and 359 we identified 130 and 7 YSOs, respectively.
   We calculate kinematic ``traceback ages'' for YSOs in these clusters, which we compare to isochronal ages estimated using several sets of stellar evolution models. 
   We find for NGC 2264 N \& S that kinematic ages are generally smaller than their isochronal ages, which may indicate these systems remained bound for a few Myr before their current state of expansion. On the other hand, kinematic ages for Collinder 95 are often significantly larger than isochronal ages, which implies many of these YSOs did not originate from a central, dense region, leading to overestimated kinematic ages.}
   {We conclude that NGC 2264 N \& S clusters likely formed as initially bound and compact systems, but have been gradually evaporating as cluster members become unbound, forming halos of unbound YSOs surrounding the cluster cores. We conclude that Collinder 95 likely formed initially sparse and substructured and has been dispersing since gas expulsion. }


   \maketitle
%

\section{Introduction}

The vast majority of stars are born in stellar clusters \citep[e.g.,][]{lada03,gutermuth09} and so understanding the formation and evolution of these clusters is critical to shedding light on the star formation process. However, age estimates for these clusters produced by fitting stellar evolutionary models to observational data suffer from many sources of uncertainty, particularly for the low-mass PMS stars that are the majority of the population in young clusters \citep{forbidden}. 

Kinematic ages offer a model-independent alternative for estimating the age of a stellar cluster as well as providing additional insight into its evolution and history. A cluster's kinematic age, or traceback age, reflects a physical consequence of prolific star formation in young stellar clusters. In these clusters, feedback from young massive stars expels surrounding molecular gas, causing the cluster to lose most of its binding mass \citep[e.g.,][]{goodwin}. This produces an expanding halo of young stars around the cluster center, an expansion trend reflected in some simulations of young clusters \citep[e.g.,][]{farias} and also in many recent observational works \citep{kuhn19,Armstrong22,guilherme23,wright24,lambda_ori} which are demonstrating that the majority of nearby young clusters exhibit such signatures of expansion. These expansion signatures can also be used to constrain the age of the cluster \citep{miret-roig24,lambda_ori}. By identifying young stellar objects (YSOs) moving outwards from the cluster center and calculating the time required for them to travel to their current positions from their assumed initial configuration, we are able to calculate a `kinematic' or traceback age of a stellar cluster. This kinematic age analysis has been greatly enabled by the advent of \textit{Gaia}'s high precision position and proper motion measurements \citep{gaia}, with which we can probe the internal kinematics of stellar clusters.

\citet{lambda_ori} investigated the internal kinematics of the $\lambda$ Ori cluster using high probability members from the \citet{cantat-gaudin20} catalog, updated with Gaia DR3 astrometry, and cross-matched with the radial velocity catalog of \citet{tsantaki22}. Using both the $Q$-parameter \citep{cartwright04} and Angular Dispersion parameter \citep{dario14}, they found evidence that the cluster contains significant substructure outside the smooth central cluster core. \citet{lambda_ori} found strong evidence for asymmetric expansion in the $\lambda$ Ori cluster, and determined the direction at which the rate of expansion is at a maximum. They also inverted the maximum rate of expansion of $0.144^{+0.003}_{-0.003}$ km/s/pc to give an expansion timescale of $6.944^{+0.148}_{-0.142}\:$Myr, which they compared to other kinematic age methods applied to this cluster \citep{squicciarini21,quintana22,pelkonen24} and literature age estimates \citep{kounkel18,zari19,cao22}. \citet{lambda_ori} also found significant asymmetry in the velocity dispersions and signatures of cluster rotation in the plane-of-sky. Putting all of these results together, they concluded that the $\lambda$ Ori cluster likely formed in a sparse, substructured configuration, and is not simply the dispersing remnant of an initially bound, monolithic cluster which began to expand after the dispersal of its parent molecular cloud. The asymmetric kinematic signatures, and the discovery of a group of candidate ejected cluster members in particular, suggest a more complex dynamical history for the $\lambda$ Ori cluster.

We seek to estimate the kinematic ages for clusters Collinder 95, NGC 2264, and Collinder 359, building on methods developed by \citet{lambda_ori} (Paper I). We establish a robust list of probable YSOs in these clusters through a combination of spectroscopic youth indicators and \textit{Gaia} DR3 variability YSO flag \citep{gaia_var}. 
We create a robust list of YSOs and, using their precise Gaia astrometry, we calculate the kinematic ages of clusters Collinder 95 and NGC 2264, omitting Collinder 359 due to an insufficient amount of identified YSOs ($N<10$). 
We then compare these kinematic age estimates to traditional isochronal age estimates using both the \citet{baraffe} and PARSEC models \citep{parsec}, and infer the likely formation history  and subsequent evolution of these clusters. 

\section{Data} 

\begin{figure*} 
	\centering
	\includegraphics[width=0.47\textwidth]{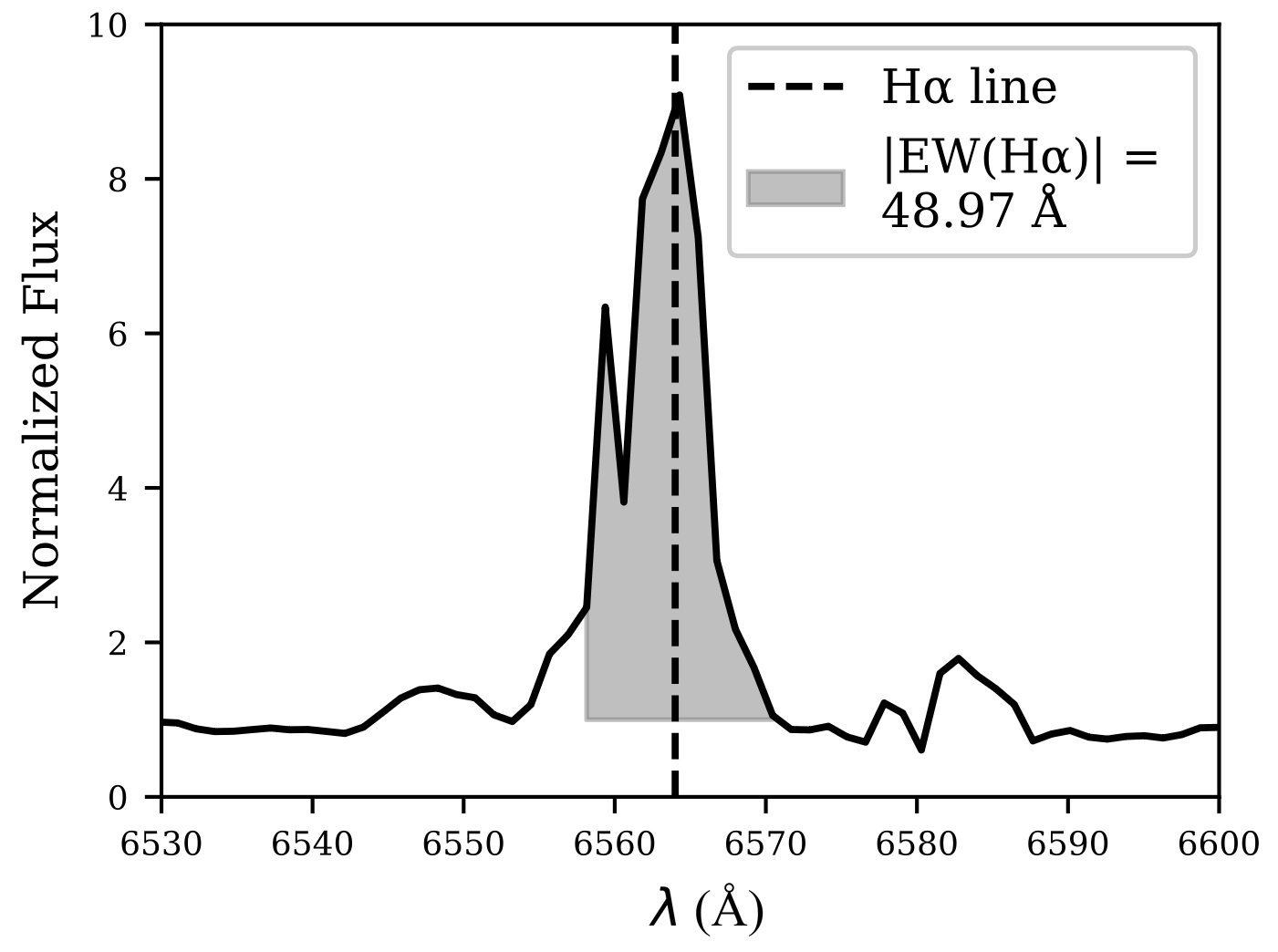}
	\includegraphics[width=0.48\textwidth]{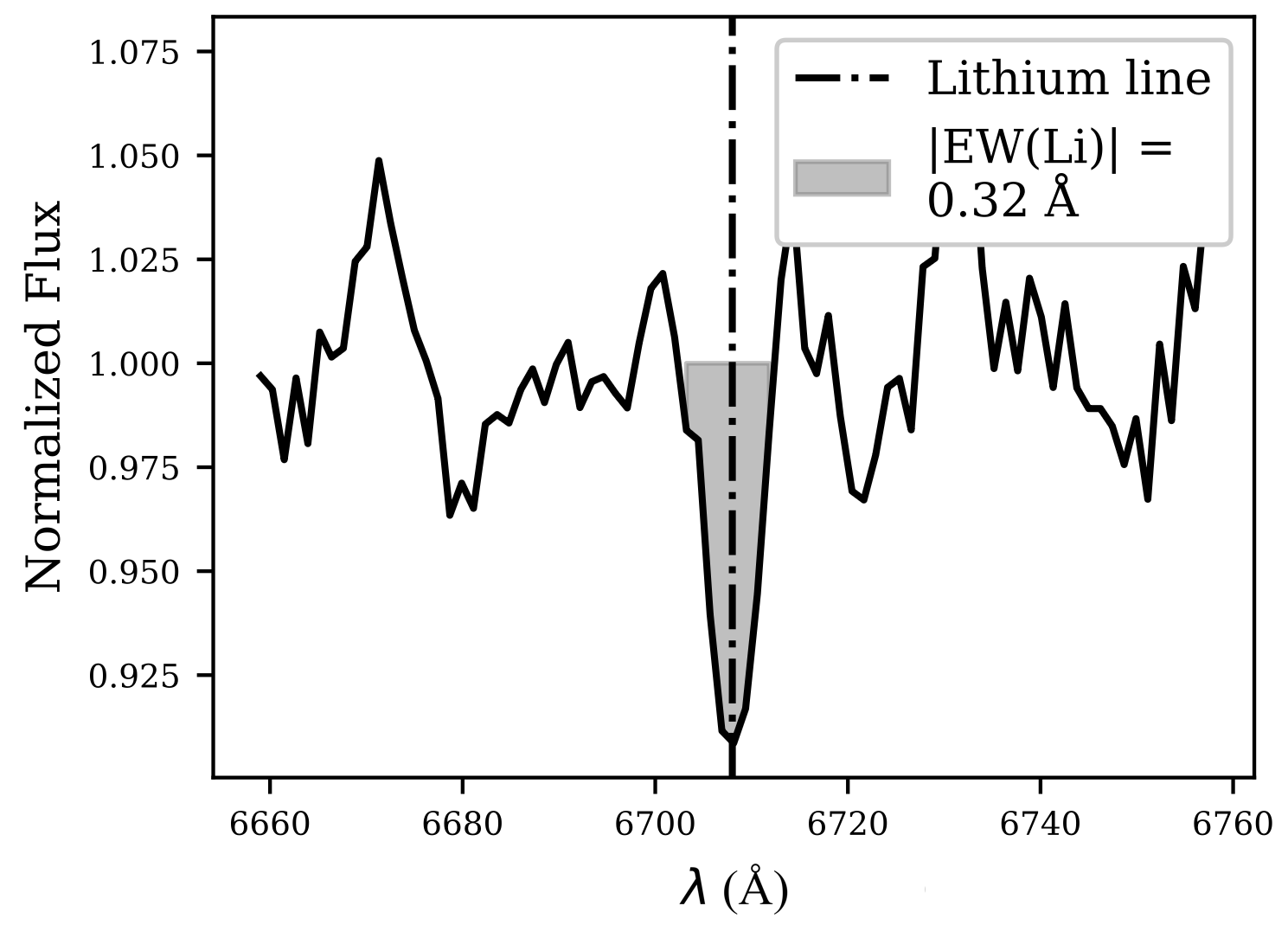}
    \caption{\emph{(a) Left:} Example H$\alpha$ spectrum of YSO candidate (\textit{Gaia} Source ID: 3326702657541965952) from NGC~2264,
    with the source selected as a YSO due to its strong H$\alpha$ line, i.e., with an EW(H$\alpha$) $= -48.97${\AA}. \emph{(b) Right:} Example $6708${\AA} Li spectrum of YSO candidate (\textit{Gaia} Source ID: 3326714782234731520) from NGC~2264, with this source selected as a YSO due to its strong Li absorption line, i.e., with EW(Li) $= 0.32${\AA}. 
    }
    \label{Fig:EW_examples}
\end{figure*}

In our analysis, we combined existing kinematic data from \textit{Gaia} DR3 with new optical spectroscopic observations of low-mass PMS candidates to create a lists of YSO candidates for each target young cluster. We then performed a kinematic age analysis on these samples.

\subsection{Target selection}
\label{target_selection}

The target clusters were chosen for their relatively young estimated isochronal ages, making them prime targets for an in-depth and insightful kinematic age analysis. They were also selected due to their relatively close line-of-sight distance ($<1\:$kpc), allowing spectra to be obtained for the more numerous lower-mass cluster members. The majority of the known YSO populations of these clusters are also within a $1^\circ$ diameter on the sky, meaning we could observe both the central cluster core and sparse halo within the MMT field-of-view.


NGC 2264 is a nearby massive young cluster ($\sim760\:$pc, $\sim3\:$Myr; \citealt{venuti18}), second only to the Orion Nebula Cluster in terms of proximity, mass and well-defined pre-main sequence (PMS) population \citep{dahm08}, where star formation is still ongoing within its parental cloud. PMS members of NGC~2264 have been identified with X-ray observations \citep[e.g.,][]{flaccomio06,guarcello17}, photometric variability \citep{cody14}, H$\alpha$ emission \citep{dahm05} and other spectroscopic youth indicators (GES; \citealt{venuti18}), belonging to a total population of $>$700 \citep{dahm05}, exhibiting hierarchical structure with multiple subclusters, in particular, two embedded star-forming clumps (NGC~2264~N / S~Mon and NGC~2264~S / Cone Nebula Cluster) are surrounded by a halo of older sparsely distributed YSOs \citep{sung08}, indicating an extended period of sequential star formation with a complex dynamical history \citep{nony,flaccomio}. 

In addition, 2.5 degrees west of NGC~2264 lies the relatively understudied young cluster Collinder 95, of similar age and distance \citep{cantat-gaudin20}. Both of these young clusters are associated with an extensive molecular cloud complex which has been found to contain a sparse, widely distributed population of YSOs \citep[Mon OB1;][]{rapson14}, as well as many protostars, again suggesting multiple epochs of star formation in this region. 

Collinder 359 was chosen for similar reasons, being a young open cluster at close distance of 250~pc \citep{coll359}. While young, at $\sim 30\:$Myr, Collinder 359 is estimated to be significantly older than NGC 2264 and Collinder 95, which aligns with us ultimately identifying the fewest number of YSOs in Collinder 359 out of all three clusters.

Candidates YSO members of these clusters were selected by fitting low-mass PMS \citet{baraffe} isochrones to \textit{Gaia} estimates of BP-RP color and G-band magnitude values for stars across all three cluster fields. YSO candidates located above younger isochrones in the color-magnitude diagram were given greater target priorities for observations.  
Targets were selected in the Gmag range 14-17.5 for NGC 2264 N \& S and Coll 95 and range 13-17 for Coll 359 with the goal to find low-mass ($\sim1.0 - 0.3 M_\odot$) cluster members
and to keep fibre cross-talk in the observations to a minimum.
Targets were also selected within parallax range $2.2 - 1.0\:$mas to be in the right distance range $\sim700 \pm 300\:$pc for NGC~2264 and Coll~95, with a parallax $> 1.25\:$mas for Coll 359, and selected with $RUWE < 1.4$ to prioritise cluster members with reliable Gaia astrometry. 

MMT-HectoSpec has a 1 degree diameter field-of-view and 300 fibres, which sets the limit for number of targets we can observe per field. The final list of targets are selected after running a fibre-configuration algorithm in 'xfitfibs', which calculates the optimal arrangement of fibres on targets without fibres crossing and with sufficient space between them (20 arcseconds), whilst weighting target priorities. This means that targets in denser regions are more difficult to allocate fibres to, biasing selection away from targets in the core of a cluster, but sampling the sparse cluster halo well.

Central coordinates in RA, Dec for observed fields were (100.2, 9.7) for NGC~2264, including both N \& S subclusters in the same FOV, (97.8, 10.0) for Coll 95 and  (270.4, 3.1) for Coll 359. As well as $\sim250$ science targets within a 1 degree diameter of these central positions, 30 - 40 HectoSpec fibres were allocated to empty sky for sky spectrum subtraction. Candidate guide stars were also selected from Gaia in the Gmag range 12 - 15 and are used at the edge of the HectoSpec field, not on the surface where the fibres are positioned.

\subsection{Data reduction}
\label{data_reduction}

Using these isochronally-selected YSO candidates, we observed $>$600 stars in four nearby young clusters (NGC 2264 N, NGC 2264 S, Collinder 95, and Collinder 359) using the MMT / Hectospec spectrograph. We observed using the 270 line mm$^{-1}$ grating in the optical range ($\sim 3700-9150${\AA}) in order to identify and flag broad H$\alpha$ emission lines and lithium absorption lines—features indicative of low-mass YSOs which are still in their PMS evolution \citep{soder}. 

Collinder 95 was observed on Feb 27th 2022 (3 x 40 minute exposures), NGC 2264 on 1st March (3 x 40 minute exposures), Coll 359 on April 4th (4 x 40 minute exposures). 

For each night of observations bias, dark, domeflat and sky flat fields were taken as well as the target fields, and data reduction including sky subtraction and cosmic ray rejection was performed using the HSRED v2.0 pipeline.

Ultimately, there were 202 combined spectra obtained in the Collinder 95 cluster with a median SNR of 36.07, 200 in the NGC 2264 N \& S clusters with a median SNR of 30.5, and 212 in the Collinder 359 cluster with a median SNR of 53.1.

\begin{figure} [htbp]
	\includegraphics[width=1\columnwidth]{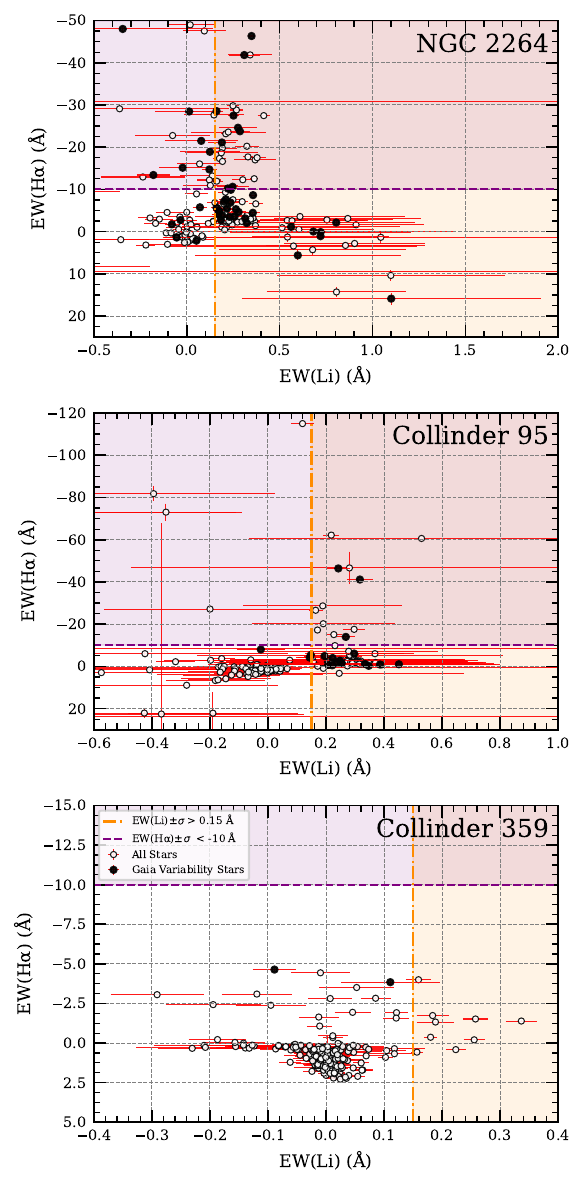}
    \caption{Scatter plots of H$\alpha$ equivalent width vs lithium equivalent width for all the stellar spectra in our samples of clusters NGC 2264 (N \& S), Collinder 95, and Collinder 359. If a star has an EW(H$\alpha$)$-\sigma< -10${\AA} (above the horizontal blue dashed line), or an EW(Li)$-\sigma>0.15${\AA} (right of the vertical green dot-dashed line), it is flagged as a YSO. \textit{Gaia} variability catalog stars are additionally marked with a closed circle.}
    \label{Fig:EW_selection}
\end{figure}

\begin{table*}[htp]
\footnotesize

\begin{tabular}{@{} *{7}{>{\centering\arraybackslash}p{2.2cm}} @{} } 
\toprule

Star Cluster & Spectroscopic Targets Observed &  $|$EW(H$\alpha$)$|$ $>10${\AA} & EW(Li) $>0.15${\AA} & Spectroscopically Flagged YSOs & \textit{Gaia} variability YSOs \citep{gaia_var} & Total YSOs \\  
\midrule \addlinespace
 NGC 2264 N\&S & 199 & 64 & 85 & 124 & 268 & 354 \\
 Collinder 95 & 202 & 19 & 23 & 35 & 104 & 130 \\
 Collinder 359 & 212 & 0 & 7 & 7 & 2 & 7 \\

\bottomrule
\end{tabular}
\caption{Number of identified YSOs in the clusters (NGC 2264 N\&S, Collinder 95, and Collinder 359), separated by method (H$\alpha$ emission line equivalent width, lithium absorption line equivalent width, and {\it Gaia} variability YSO flag). For YSOs identified by \textit{Gaia} YSO variability, we only include stars with a G-band mean flux divided by its error greater than $50$, an integrated RP mean flux divided by its error greater than $20$, an integrated BP mean flux divided by its error of greater than $20$, and greater than $5$ visibility periods.}
    \label{tab:yso_numbers}
\end{table*}

\begin{figure} 
	\centering
	\includegraphics[width=0.8\columnwidth]{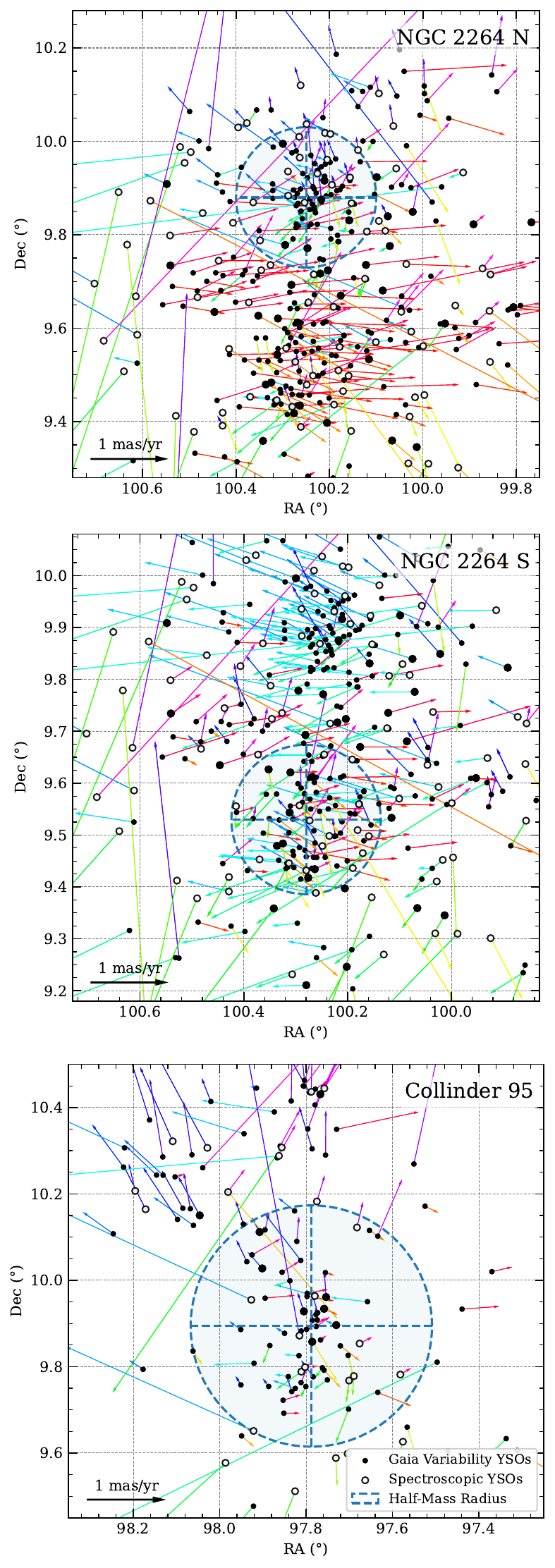}
    \caption{Proper motion maps of YSOs in NGC 2264 N, NGC 2264 S, and Collinder 95. Stars flagged as YSOs via our spectroscopic analysis are marked with an open circle, YSOs flagged through the \textit{Gaia} variability catalog are marked with smaller black point. (Stars flagged by both indicators are denoted through a filled in circle / larger black point.) The proper motion of each YSO is shown with an arrow which is color-coded according to the direction of motion. The average motion of each cluster or subcluster has been subtracted out (see these values in Table \ref{tab:cluster_info}), allowing us to observe the plane of sky motion of each YSO within the cluster frame. The center of the cluster is denoted by the intersection point of the two blue dotted lines. The dotted blue circle denotes the half-mass radius of the identified YSOs in each cluster.}
    \label{Fig:colormap_grid}
\end{figure}

\begin{figure} 
	\centering
	\includegraphics[width=0.9\columnwidth]{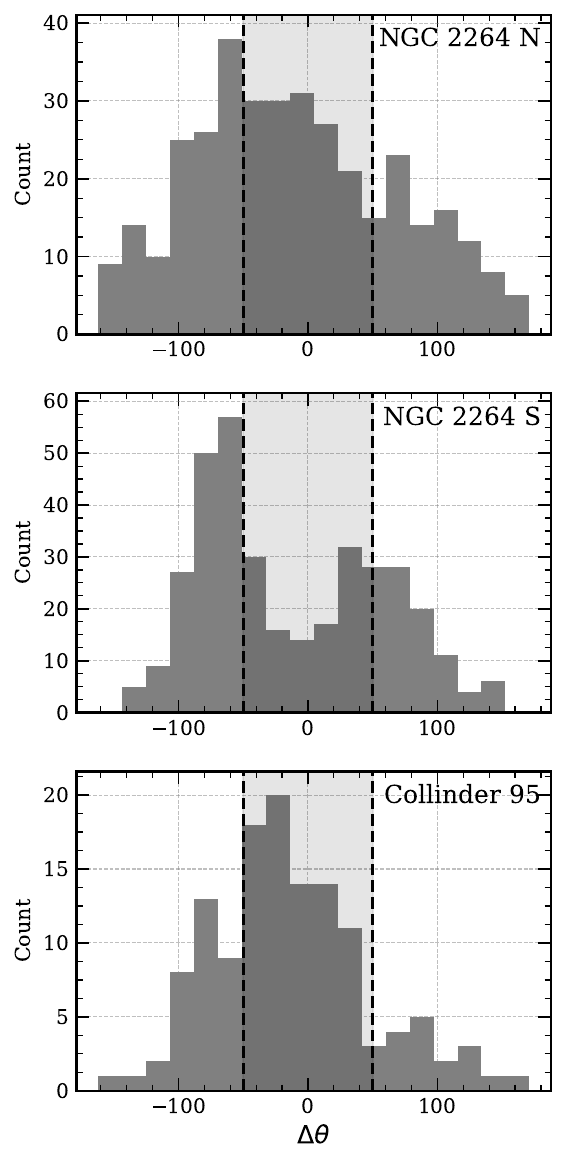}
    \caption{Expansion direction histograms of YSOs in NGC 2264 N, NGC 2264 S, and Collinder 95, where the measured angle, $\Delta\theta$, is the difference in a YSO's direction of motion from pure radial outward motion from the cluster center. 
    We observe that there is a large proportion of stars near zero ($\pm 50$\textdegree) for NGC 2264 N and Collinder 95, indicating that there is significant outward expansion of YSOs from their cluster centers. We do not observe this same trend for NGC 2264 S. The shaded region shows a range $\pm 50^\circ$, defining a subset of YSOs with strongest radial expansion. 
    }
    \label{Fig:all_angles}
\end{figure}

\section{Young Stellar Object Identification}

There are a few spectral features which can be used to identify young PMS stars. A strong H$\alpha$ emission line ($6564${\AA}), caused by residual accretion and/or enhanced chromospheric activity, appears in spectra for stars $< 10\:$Myr (Classical T Tauris; CTTs), whereas a strong lithium absorption line ($6708${\AA}) is visible in spectra for PMS stars $< 20\:$Myr \citep{soder}. Once the base of the convection zone in PMS stars reaches $3 \times 10^6\:$K, Li is rapidly depleted, on a timescale dependent on their spectral type, which makes the presence of the Li line a strong discriminator between PMS stars and older field stars.

We searched for prominent H$\alpha$ emission and lithium absorption lines in the spectra of our observed candidate cluster members, quantified by their equivalent width (EW) values. We identified YSO candidates if they have $|$EW(H$\alpha$)$|$ $> 10${\AA} or EW(Li) $> 0.15${\AA}, as these are established thresholds for these spectroscopic indicators of youth \citep[e.g.,][]{h_alpha,lithium}. Two example YSO candidate spectra are shown in Figure~\ref{Fig:EW_examples}. 
In Figure \ref{Fig:EW_selection} we plot EW(H$\alpha$) versus EW(Li) for all the stars observed in our target clusters with the YSO criteria thresholds indicated as dashed lines. 


We find $124$ YSOs in NGC 2264, $35$ in Collinder 95, and $7$ in Collinder 359 using these spectroscopic indicators of youth. To complement this sample of probable YSOs, overlapping with the same field of view we add in YSOs flagged by \textit{Gaia} DR3 variability \citep{gaia_var}: 
$268$ YSOs in NGC 2264, $104$ in Collinder 95, and $2$ in Collinder 359. 
We also quality filter in this step, rejecting stars with a G-band mean flux divided by its error of less than $50$, an integrated RP mean flux divided by its error of less than $20$, an integrated BP mean flux divided by its error of less than $20$, or with $5$ or fewer visibility periods. 

Between these two methods of YSO identification—spectroscopic and the \textit{Gaia} DR3 variability catalog—some stars were flagged using both methods. Stars which were flagged by both methods numbered $38$ in NGC 2264, $9$ in Collinder 95, and $2$ in Collinder 359. A star was listed as a YSO in the final tally if it was flagged by either method. 

A detailed breakdown of the number of YSOs identified via each method in each cluster is available in Table \ref{tab:yso_numbers}. Using these two methods, we generate a list of robust YSO candidates, ultimately selecting a total of $354$ YSOs in NGC 2264, $130$ in Collinder 95, and $7$ in Collinder 359.

\section{Kinematic Analysis}

We then use this population of YSOs to conduct a kinematic analysis of the star clusters, excluding Collinder 359 as its sample contained too few identified YSOs to produce statistically significant results. Therefore, we only conduct kinematic analyses of the YSOs in the star clusters NGC 2264 N, NGC~2264~S and Collinder 95. 

\subsection{Proper motion orientation}

We search for signs of expansion in the plane-of-sky among our YSOs by measuring the angle between each star's cluster frame and virtual expansion corrected proper motion vector and the vector between the cluster center and the star's current location. 
The cluster frame motion is found by subtracting out the mean cluster proper motion from that of each YSO. 

The adopted mean proper motions, radial velocities, center locations and half-mass radii of each cluster are listed in Table \ref{tab:cluster_info}.
We use center coordinates from existing literature (see Table \ref{tab:cluster_info}), while YSO half-mass radii derived from our identified YSO population. In particular, we check recent cluster catalogs with membership based on Gaia \citep[e.g.,][]{cantat-gaudin20,hunt24} for cluster parameters. For Collinder 95 we take central coordinates, proper motions and distance from \citet{cantat-gaudin20}. However, for NGC~2264, these catalogs do not distinguish between the two apparent subclusters, NGC~2264 N \& S. Instead, we use cluster parameters for NGC 2264 N \& S from \citet{aanda}, except for the central proper motions of NGC~2264~S, which in \citet{aanda} are given as identical to NGC~2264~N. For the central proper motions of NGC~2264~S we take the median proper motion of our confirmed YSOs with Dec. $< 9.7^\circ$.

The cluster frame proper motions of the YSOs in each cluster are visualized in Figure \ref{Fig:colormap_grid}. These plots show arrows denoting each star's plane of sky motion with the direction of this motion color-coded according to the color wheel in the top left corner. \textit{Gaia} YSOs and spectroscopically identified YSOs are marked differently on the plots for clarity. In Figure \ref{Fig:colormap_grid}, the distinct proper motions of NGC 2264's two subclusters become clear through the plotted vectors. We note that YSOs outside the half-mass radii, indicated using dashed circles, appear to be preferentially moving away from the cluster centers, which is particularly apparent for NGC~2264~N and Collinder 95. 

We plot the angle between each YSO's proper motion and the radial direction from cluster center to the YSO, $\Delta\theta$, producing the distributions shown in Figure \ref{Fig:all_angles} for all three clusters. If there exists an expanding halo of YSOs in these clusters, we would expect a substantial peak around zero degrees. We observe this trend for NGC~2264~N and Collinder 95, but not for NGC~2264~S.

\subsection{Tangential velocities}


Virtual expansion is the projection effect in plane-of-sky motions created when a population of stars is receding or approaching in the line-of-sight. We corrected for this effect using equations provided in \citet{virtual_expansion} using each cluster's bulk radial velocity as provided in the \citet{cantat-gaudin20} catalog. We used an MCMC approach to sample uncertainties from the posterior distribution for the final corrected tangential velocities $v_{l}$ and $v_{b}$, similar to the approach used by \citet{lambda_ori}. We use these velocities in our subsequent kinematic analysis.

\subsection{Average expansion velocities}
\label{expansion velocities}

Another indicator for cluster expansion is the median cluster expansion velocity $\bar v_{\rm out}$, i.e., the average of their radial tangential velocities.
As described in \citet{lambda_ori}, we measure the cluster expansion velocity by taking the median expansion velocity component of all cluster members for 
$10^6$ iterations with additional uncertainties randomly sampled from the observed velocity uncertainties. The uncertainties on the cluster expansion velocity are then taken as the 16th and 84th percentile values of the posterior distribution. Note, for the purpose of allocating YSOs to either NGC 2264 N or S clusters, we only include YSOs within a $0.175^\circ$ radius of either cluster's central position when calculating $\bar v_{\rm out}$. Expansion velocities and uncertainties for each cluster are presented in Table~\ref{tab:cluster_info}.

We find that NGC 2264 N has $\bar v_{\rm out} = 0.45^{+0.11}_{-0.11}$ km/s, NGC 2264 S has $\bar v_{\rm out} = 0.23^{+0.10}_{-0.10}$ km/s and Collinder 95 has $\bar v_{\rm out} = 0.51^{+0.08}_{-0.08}$ km/s. These positive median velocities provide evidence of 4$\sigma$, 2$\sigma$ and 6$\sigma$ significance, respectively, that these clusters are expanding. 

These velocities are lower than the expansion velocity found for $\lambda$ Ori of $\bar v_{\rm out} = 0.71^{+0.02}_{-0.02}$ km/s by \citet{lambda_ori}, but are comparable to the expansion velocities found for $\delta$ Sco and $\sigma$ Sco sub-regions of Upper Scorpius by \citet{armstrong25}, and sit in the middle of the range of expansion velocities calculated by \citet{kuhn19,wright24} for various nearby young clusters. In particular, \citet{kuhn19} measured expansion velocities of $\bar v_{\rm out} = 0.39 \pm 0.15$ km/s for S Mon (NGC 2264 N) and $\bar v_{\rm out} = 0.36 \pm 0.40$ km/s for IRS 1 (NGC 2264 S; but further divided into IRS 1 and IRS 2), while \citet{wright24} measured expansion velocities of $\bar v_{\rm out} = 0.40^{+0.01}_{-0.06}$ km/s for S Mon (NGC 2264 N) and $\bar v_{\rm out} = 0.23^{+0.03}_{-0.07}$ km/s for the Spokes cluster (NGC 2264 S). Our expansion velocity for NGC 2264 N is in close agreement with those from both \citet{kuhn19} and \citet{wright24}, while our expansion velocity for NGC 2254 S agrees well with that from \citet{wright24}, but the further division of NGC 2264 S into IRS 1 and IRS 2 groups by \citet{kuhn19} makes it difficult to make a direct comparison with their results.

\subsection{Areal number density profiles and cluster core sizes}
\label{DensityProfiles}

\begin{figure}
    \centering
    \includegraphics[width=240pt]{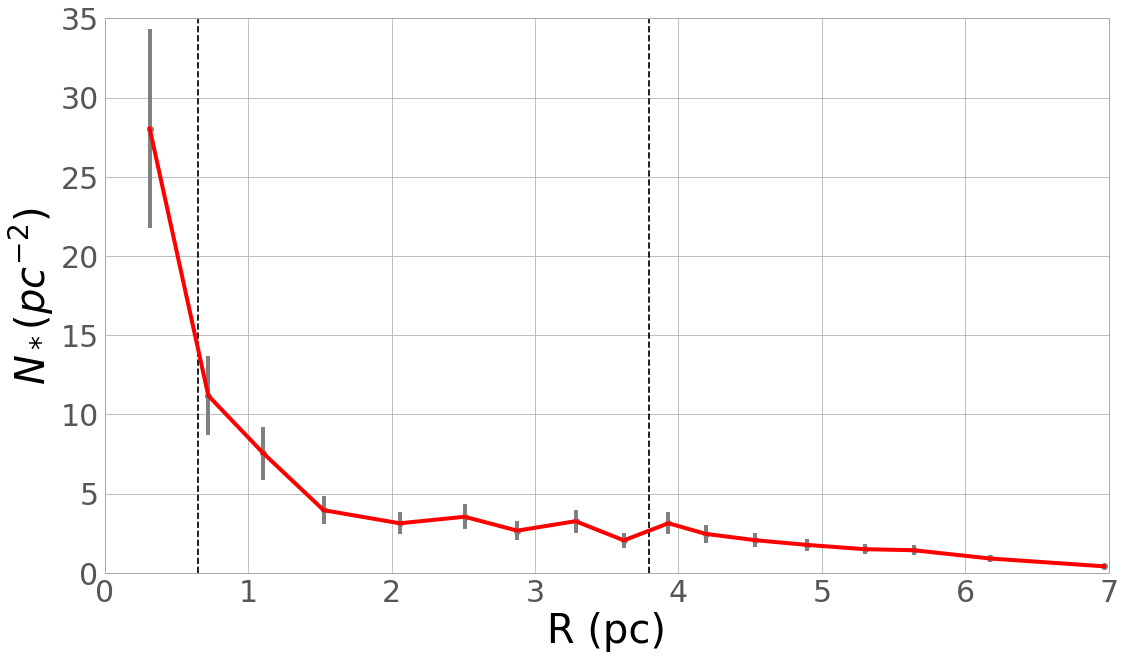}
    \includegraphics[width=240pt]{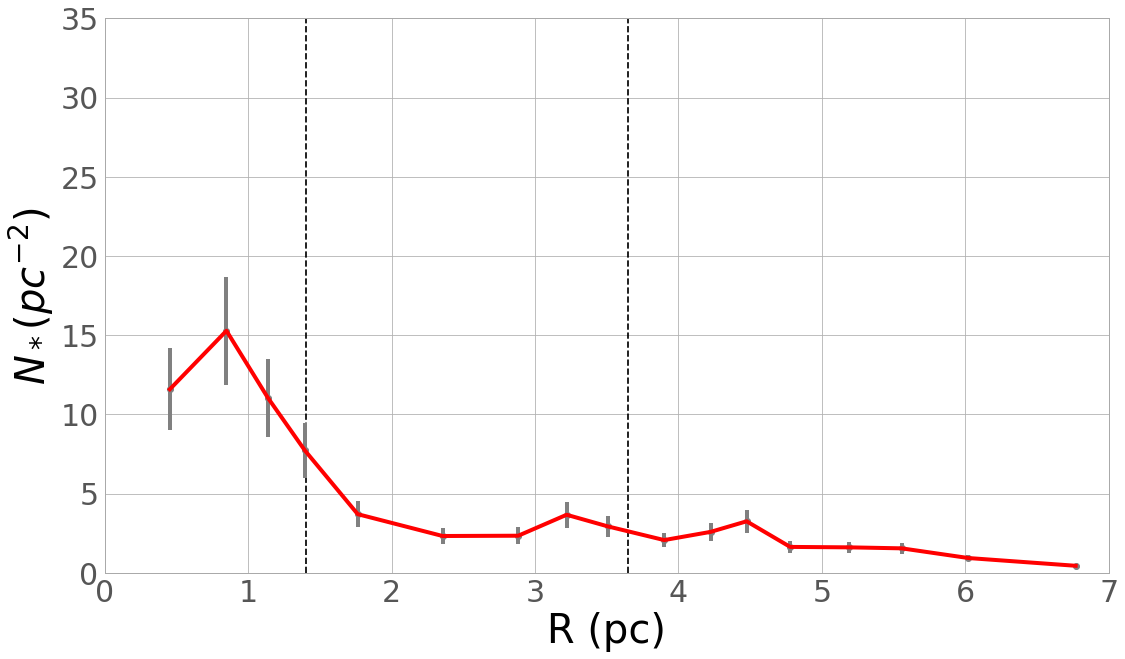}
    \includegraphics[width=240pt]{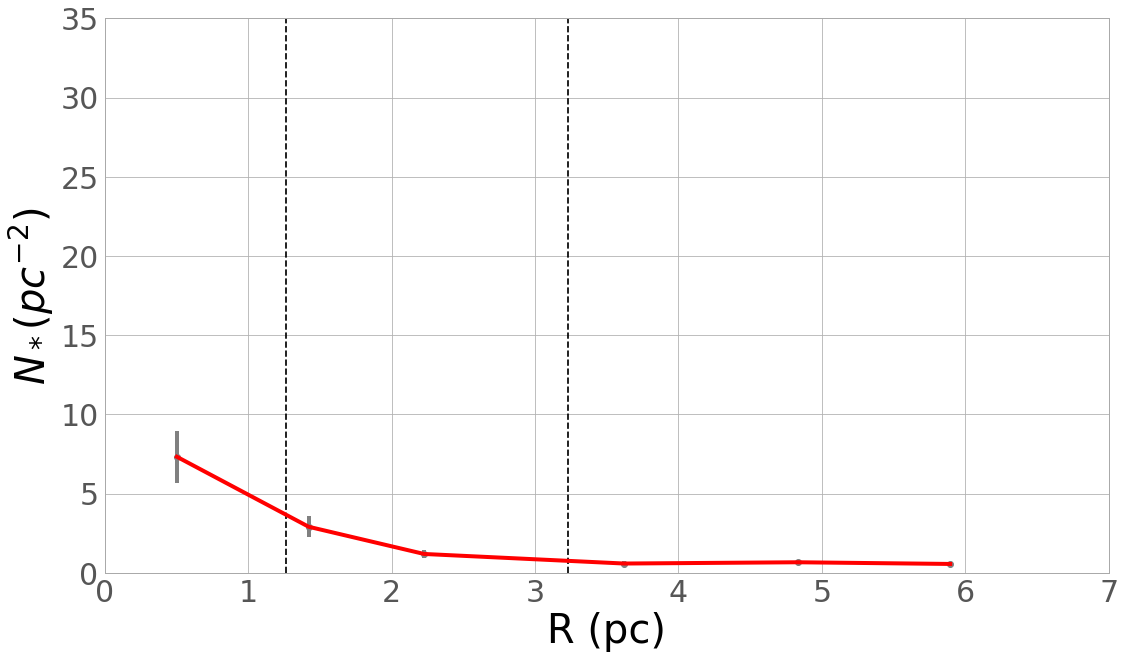}
    \caption{Radial profiles of areal stellar number density profiles of NGC~2264~N (\textit{Top}), NGC~2264~S (\textit{Middle}) and Collinder 95 (\textit{Bottom}). Vertical dashed lines indicate the core radii ($r_c$), as defined in Section~\ref{DensityProfiles}, and half-mass radii ($r_{50}$).}
    \label{fig:RadialProfiles}
\end{figure}

We investigate the radial density profiles of the clusters and define the radii of the cluster cores using the same approach as \citet{lambda_ori}. We bin cluster members by increasing radial distance from the cluster center, $R$.
We calculate the areal number density, $N_*$, of each bin.
The resulting radial areal number density profiles are shown in Fig.~\ref{fig:RadialProfiles}.

We define the core radius of a cluster, $r_c$, as the radius at which the density is half the peak density, which is $r_c=0.65\:$pc for NGC~2264~N, 1.40~pc for NGC~2264~S and 1.27~pc for Collinder 95. We find that 24, 72 and 26 cluster members are located within the core radii of NGC~2264~N, NGC~2264~S and Collinder 95, respectively.

The areal density profiles of NGC~2264~N and Collinder 95 smoothly decrease with increasing radius, while the profile of NGC 2264 S has its peak density offset from the central coordinates. This is likely due to the substructure that \citet{kuhn19} defined as the IRS 1 and IRS 2 subgroups of the cluster, both of which are contained within our resulting core radius for NGC~2264~S.

\begin{table*}[htp]
\centering
\begin{tabular}{|p{6.8cm}|p{2.8cm}|p{2.8cm}|p{2.8cm}| }
\toprule
   & NGC 2264 N & NGC 2264 S & Collinder 95 \\
\midrule
Central RA, Dec (\textdegree) & (100.25, 9.88)\TblrNote{1} & (100.28, 9.53)\TblrNote{2} & (97.788, 9.894)\TblrNote{3} \\
Central $\mu_{\alpha}$ (mas/yr) & $-1.72\substack{+0.06\\ -0.06}$\TblrNote{2} & $-2.18\substack{+0.66\\ -0.42}$\TblrNote{4} & $-2.26\substack{+0.35\\ -0.35}$\TblrNote{3} \\
Central $\mu_{\delta}$ (mas/yr) & $-3.71\substack{+0.06\\ -0.06}$\TblrNote{2} & $-3.75\substack{+0.27\\ -0.40}$\TblrNote{4} & $-5.16\substack{+0.31\\ -0.31}$\TblrNote{3} \\
Distance (pc) & $719\substack{+22\\ -21}$\,\TblrNote{3}  & $722\substack{+22\\ -21}$\,\TblrNote{3} & $659.0\substack{+86.7\\ -24.1}$\,\TblrNote{3} \\
Expansion velocity $\bar v_{\rm out}$ (km/s) & $0.45^{+0.11}_{-0.11}$ & $0.23^{+0.10}_{-0.10}$ & $0.51^{+0.08}_{-0.08}$ \\
Core radius $r_c$ (pc) & 0.65 & 1.40 & 1.27 \\
No. YSOs within $r_c$ & 24 & 72 & 26 \\
Half-mass radius $r_{50}$ (pc) & 3.83\TblrNote{5} & 3.70\TblrNote{5} & 3.69\TblrNote{5} \\
Velocity dispersion $\sigma_{v_l}(r_c)$ (km/s) & $0.96^{+0.25}_{-0.19}$ & $1.29^{+0.17}_{-0.14}$ & $5.39^{+1.11}_{-0.86}$ \\
Velocity dispersion $\sigma_{v_b}(r_c)$ (km/s) & $0.65^{+0.25}_{-0.18}$ & $1.77^{+0.23}_{-0.19}$ & $4.27^{+0.96}_{-0.70}$ \\
Velocity dispersion $\sigma_{v_l}(r_{50})$ (km/s) & $2.56^{+0.22}_{-0.19}$ & $3.15^{+0.33}_{-0.30}$ & $3.51^{+0.44}_{-0.37}$ \\
Velocity dispersion $\sigma_{v_b}(r_{50})$ (km/s) & $3.60^{+0.26}_{-0.23}$ & $2.27^{+0.18}_{-0.15}$ & $3.51^{+0.45}_{-0.37}$ \\
Crossing time $\tau_{\rm cross}(r_c)$ (Myr) & $0.79^{+0.18}_{-0.13}$ & $0.91^{+0.08}_{-0.07}$ & $0.26^{+0.04}_{-0.03}$ \\
Crossing time $\tau_{\rm cross}(r_{50})$ (Myr) & $1.23^{+0.07}_{-0.06}$ & $1.34^{+0.09}_{-0.08}$ & $0.92^{+0.08}_{-0.07}$ \\
No. YSOs with closest approach $<r_c$ & 24 & 29 & 21 \\
No. YSOs with closest approach $-\sigma<r_c$ & 63 & 50 & 49 \\
No. YSOs with closest approach $-2\sigma<r_c$ & 113 & 69 & 60 \\
No. YSOs with closest approach $-3\sigma<r_c$ & 132 & 81 & 61 \\
\bottomrule
\end{tabular}
\caption{Cluster physical properties. See the text for a discussion of how these quantities were derived. \TblrNote{1} Cluster center value taken from \citet{aanda} and adjusted (+0.05\textdegree, +0) to center accurately on the cluster core of YSOs. \TblrNote{2} Values taken from \citet{aanda}. \TblrNote{3} Values taken from \citet{cantat}. \TblrNote{4} NGC 2264 S central proper motion value calculated by taking the median proper motion value of all YSOs located below a declination of $9.7$\textdegree. \TblrNote{5} Half-mass radii of each cluster calculated by taking the median distance from cluster center of all member YSOs. }
  \label{tab:cluster_info}
\end{table*}

\subsection{Velocity dispersions and crossing times}

We calculate velocity dispersions in each cluster using the same Bayesian inference approach described in \citet{lambda_ori}. We model velocity distributions as 2-d Gaussians with free parameters for the central velocity $\mu$ and velocity dispersion $\sigma$ in each dimension. We then add an uncertainty randomly sampled from the observed uncertainty distribution in each dimension for each star in a cluster.
We sample the posterior distribution function with MCMC, using an unbinned maximum likelihood test to compare the model and observations. As a prior we require that velocity dispersions must be $>0$ km~s$^{-1}$. We repeat for 2000 iterations with 1000 walkers, the first half of which are discarded as burn-in. We then take the median of the posterior distribution as the best fit for each free parameter and the 16th and 84th percentiles as their respective 1$\sigma$ uncertainties.

With this approach we calculate velocity dispersions for cluster members within the cluster cores and for cluster members within the cluster half-mass radii, using their virtual-expansion corrected tangential velocities ($v_l, v_b$). Using these velocity dispersions along with our previous estimates of cluster core radii and half-mass radii, we estimate the crossing times for core radii $\tau_{\rm cross}(r_c)$ and half-mass radii $\tau_{\rm cross}(r_{50})$ of each cluster. Both 1D velocity dispersions $\sigma_{v_l}$, $\sigma_{v_b}$ and crossing times are given for each cluster in Table~\ref{tab:cluster_info}, along with their respective uncertainties.

Notably, while the velocity dispersions in the cluster cores of NGC 2264 N \& S are relatively low ($<2$ km/s) and would not be unusual for gravitationally bound clusters, the velocity dispersion for the core of Collinder 95 is relatively large ($>4$ km/s). Given the sparseness of the cluster, this suggests it was born with a relatively high velocity dispersion and was gravitationally unbound from an early time (see Sect.~\ref{age_comparison}).
However, to properly assess gravitational boundedness requires assessment of the virial state of the clusters. This in turn requires estimates of each cluster's total mass, which is complicated by the incompleteness of our YSO sample.

For all three clusters, velocity dispersions of YSO members within their half-mass radii are relatively high ($>2$ km/s). In the cases of NGC 2264 N \& S, these velocity dispersions are significantly greater than those of their cluster cores, indicating that the halos of YSOs surrounding the cores of each cluster are likely not gravitationally bound and are dispersing. In Collinder 95 these velocity dispersions seem isotropic, while in NGC 2264 N \& S, velocity dispersions in the directions of Galactic longitude and latitude ($\sigma_{v,l}$,$\sigma_{v,b}$) are anisotropic at the 3$\sigma$ and 2.5$\sigma$ significance levels, respectively.




\subsection{Traceback Ages}

\begin{figure*} 
	\centering
	\includegraphics[width=\linewidth]{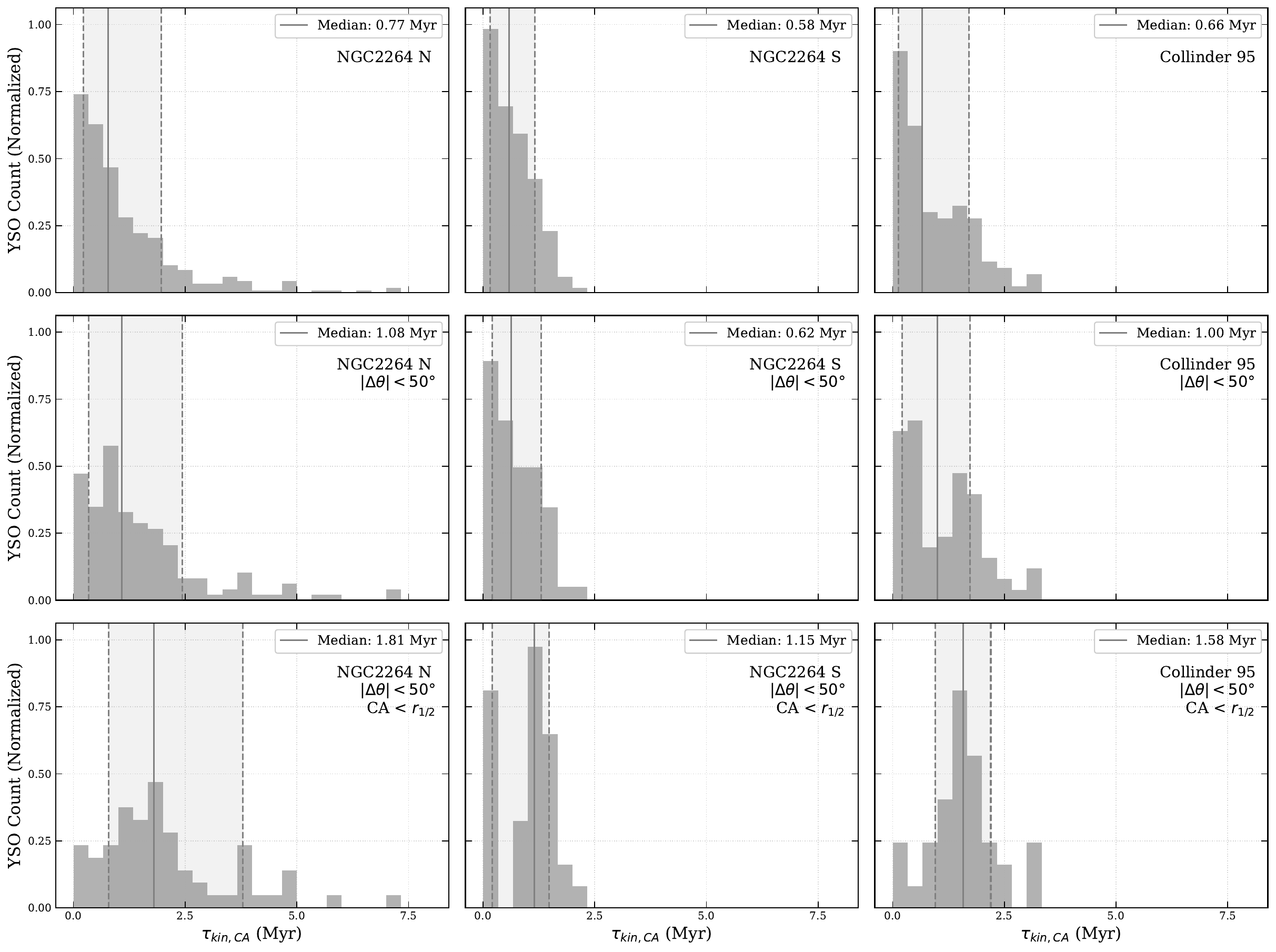}
    \caption{Traceback ages (see text) of YSOs for NGC 2264 N (left column), NGC 2264 S (middle column), and Collinder 95. Each row represents a different sub-population. In the first row, we show all identified YSOs in each cluster. In the second row, we show YSOs with $|\Delta\theta|\leq50^\circ$. In the third row, we further restrict the $|\Delta\theta|\leq50^\circ$ YSOs population by requiring them to have a current distance from cluster center greater than the half-mass radius and a distance of closest approach to the cluster center of less than the half-mass radius. The median and $\pm1\sigma$ range of the distributions are indicated.
    }
    \label{Fig:all_nine_kinematic_ages}
\end{figure*}


We calculate each cluster’s kinematic age $\tau_{\rm kin,CA}$ by ``tracing back'' each star to its moment of closest approach to the cluster center, assuming constant velocity. We repeat this calculation $N=10,000$ times, each time adding a random (normal) amount of the proper motion error to the proper motion of each star. This generates a distribution of kinematic ages for each star, which we can then use to find the estimated error bounds (16th and 84th percentiles) of each star's kinematic age. From this, we are able to calculate a kinematic age value with error estimates for each star in each stellar cluster. We then use the distribution of a cluster's YSO's kinematic ages to estimate the overall kinematic age value of a cluster. 

We start by looking only at YSOs with a particularly strong outward expansion from the cluster center, i.e., those with $\Delta\theta$ within $\pm 50$\textdegree of zero, as denoted by the dashed vertical lines in Figure \ref{Fig:all_angles}. These distributions of kinematic ages are shown in Figure \ref{Fig:all_nine_kinematic_ages}. 

For NGC 2264 N, the median kinematic age of all YSO cluster members within $\pm 50$\textdegree$ $ is $\tau_{\rm kin,CA} =1.08$ Myr with a standard deviation of $\pm1.64$~Myr. For NGC 2264 S, the median kinematic age of such YSO cluster members is $\tau_{\rm kin,CA} =0.61$~Myr with a standard deviation of $\pm0.51$~Myr. For Collinder 95, the median kinematic age of such cluster members is $\tau_{\rm kin,CA} =0.94$~Myr with a standard deviation of $\pm0.84$~Myr.  All of our kinematic age estimations are listed in Table \ref{tab:all_vals}.


When we further restrict the population we are considering to only YSOs whose current distance from cluster center is greater than the half-mass radius, shown as dashed circles in Figure \ref{Fig:colormap_grid}, and whose final closest approach distance from cluster center is less than the cluster's half-mass radius, we observe the median kinematic age values of each cluster increase slightly. The half-mass radius used here is calculated as the smallest radius that includes half of our identified YSOs in each group. We note that selection bias of spectroscopic targets away from the dense cluster cores (because of fibre crowding) might bias the half-mass radius that we calculate from our sample to moderately larger values. 
The half-mass radii for each cluster are listed in Table \ref{tab:cluster_info}.

When filtering only for YSOs whose current position is outside our half-mass radius and whose closest approach distance was inside of it, we find that: for NGC 2264 N, the median kinematic age shifts to $1.81$ Myr with a standard deviation of $1.47$~Myr; for NGC 2264 S, the median kinematic age shifts to $1.14$ Myr with a standard deviation of $0.57$~Myr; for Collinder 95, the median kinematic age becomes $1.62$ Myr with a standard deviation of $0.77$~Myr. We can observe this increase clearly in Figure \ref{Fig:all_nine_kinematic_ages}. These values are listed also in Table \ref{tab:all_vals}.

We also consider how many cluster members have their closest approach to the cluster centers within the core radii of each cluster, as defined in Section~\ref{DensityProfiles}, and which are currently located outside the cores. We list the number of YSOs per cluster with closest approach distances within the core radii with different tolerances according to the uncertainties on their closest approach distances in Table~\ref{tab:cluster_info}.


\section{Age Comparison}

\subsection{Correcting for Extinction}

\begin{figure*}
	\centering
    \includegraphics[width=0.48\textwidth]{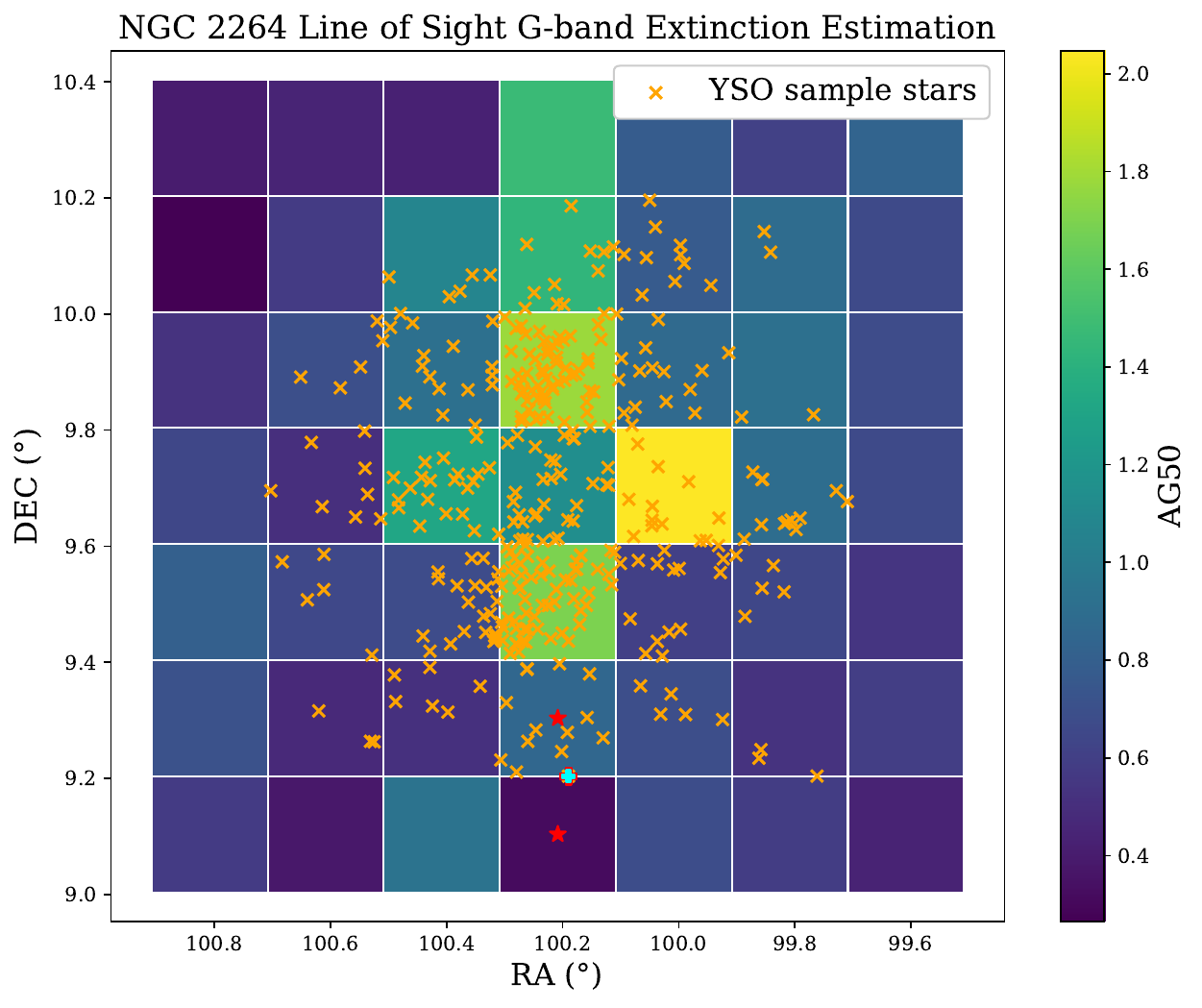}
	\includegraphics[width=0.48\textwidth]{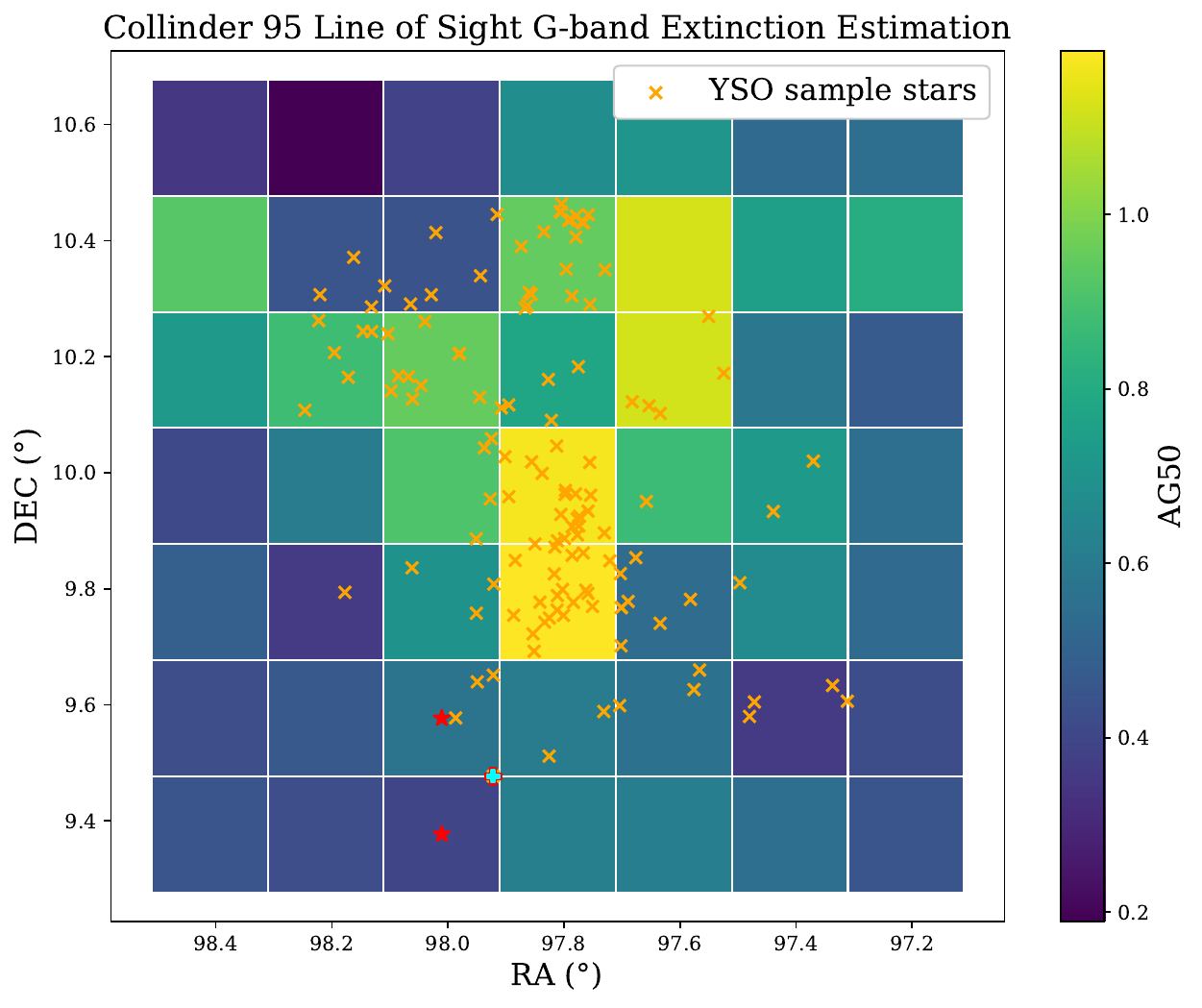}
    \caption{\emph{(a) Left:} A plot of the extinction correction estimation for NGC 2264 N \& S. The extinction values of each cell are determined by drawing from a catalog of stars' extinction values derived from the STARHORSE code and \textit{\textit{Gaia}} DR3 data \citep{starhorse}. We divide the cluster into 0.2\textdegree$ $ radial ascension by 0.2\textdegree$ $ declination cells. Then, we take the median extinction value of each cell and apply it to all the YSOs within that cell, thereby estimating each YSOs' extinction value. The YSOs are overplotted as orange crosses. The lowest number of catalog stars populating any cell is $N_{\text{min}} = 20$ for Collinder 95 and $N_{\text{min}} = 31$ for NGC 2264. \emph{(b) Right:} As (a), but for Collinder 95.}
    \label{Fig:pixel_plot}
\end{figure*}

We then seek to compare the kinematic age estimates calculated in the previous section to the isochronal ages of each cluster. To calculate these isochronal ages, we first correct our YSOs for reddening and extinction. Using extinction values from the STARHORSE catalog \citep{starhorse} and \textit{\textit{Gaia}} DR3 data, we create a local extinction ``map'' to estimate extinction values of all YSOs in our clusters. The map consists of 0.2\textdegree$ $ RA by 0.2\textdegree$ $ Dec cells as shown in Figure \ref{Fig:pixel_plot}.

We then take the median STARHORSE $A_{G}$ extinction values of all stars within each RA-Dec cell, only including stars within a certain range of parallax values. For NGC 2264, the range of parallaxes present in our YSOs is $1.018$ mas to $2.958$ mas, so when picking from the extinction catalog, we only include stars between $0.95$ mas and $3.0$ mas. For Collinder 95, the range of parallaxes present in our YSOs is $1.037$ mas to $2.818$ mas, so when picking from the extinction catalog, we only include stars between $0.95$ mas and $2.9$ mas.  

Then, we assign this median extinction value to all YSOs located within that cell. Using this method, we generate estimated extinction values for all YSOs in our sample, along with error bounds on these values (the 16th and 84th percentiles.) The minimum amount of catalog stars in each cell used to calculate these median and error bounds is $N_{\text{min}} = 31$ for NGC 2264 and $N_{\text{min}} = 20$ for Collinder 95. The extinction values estimated for each cell are visualized in Figure \ref{Fig:pixel_plot}. We follow the same approach for $A_{BP}$ \& $A_{RP}$ in order to also estimate reddening values in the Gaia photometric bands.

\subsection{Calculating Isochronal Ages}

To estimate the isochronal ages $\tau_{\rm iso}$ for cluster members, we use the \citet{baraffe}, PARSEC \citep{parsec} and SPOTS \citep{somers20} stellar isochrone models. The \citet{baraffe} model specializes in PMS stars and low-mass MS stars, with a mass range of $0.01 - 1.4 M_{\odot}$, while the PARSEC model has a mass range $>0.1 M_{\odot}$. The SPOTS models occupy a smaller mass range $0.08 - 1.3 M_{\odot}$, but are distinguished by having parameters for spot coverage fraction $F_{\rm spot}$ and spot temperature ratio $X_{\rm spot}$. Here we assume $F_{\rm spot}=0.35$ and $X_{\rm spot}=0.9$ and use the corresponding set of isochrones, as was done for the $\lambda$ Ori cluster by \citet{lambda_ori}. 

We start with the \citet{baraffe} stellar isochrones. Following the method of \citet{lambda_ori}, for each star in each cluster, we compare each individual star to a series of $\sim960$ \citet{baraffe} isochrones with ages ranging from $\sim0.5$ to $10,000$ Myr. For each star, we iterate the calculation $N = 100$ times, each time multiplying the error on the extinction value with a value randomly sampled from a normal distribution $N\sim(0,1)$ and add this to the star's median extinction value. Then, we use this new perturbed extinction value to find the isochrone closest to this star's G-band magnitude and BP-RP color. We then do the same for the G-band magnitude and G-RP color. From this series of $N=100$ iterations, we are able to generate a distribution of isochronal ages for each star for both the BP-RP and G-RP isochrones. Then, we take the median age as the isochronal age value for that star. We also take the 16th and 84th percentiles of this distribution as the upper and lower error bounds on this value. From this process, we are able to calculate the \citet{baraffe} isochronal ages for each individual star for either BP-RP and G-RP color. We follow the same approach to estimate ages using PARSEC \citep{parsec} and SPOTS \citep{somers20} isochrones.

Median isochronal ages for each cluster are given in Table~\ref{tab:all_vals}. Particularly noticeable is the clear difference between ages estimated using either BP-RP or G-RP colors. Using \citet{baraffe15} models, ages estimated using G-RP color are consistently younger than ages estimated using BP-RP color, while using \citet{parsec} models, ages estimated using G-RP color are consistently older. These discrepancies in ages estimated using different stellar evolution models and different colors have been seen for other nearby young clusters and associations in the literature \citep[e.g.,][]{ratzenbock23,lambda_ori,fajrin24}.

\subsection{Kinematic Age versus Isochronal Age }
\label{age_comparison}

In Figs. ~\ref{Fig:N2264ages},~\ref{Fig:S2264ages},~\ref{Fig:Col95ages} (\textit{top row}) we compare the isochronal age values for each set of stellar evolution models using Gaia G-RP colour with the kinematic ages of YSOs with position-velocity angles with $\pm50^\circ$. We generally expect the kinematic ages of cluster members will be lower than their isochronal ages. This is because the kinematic age calculation gives the time when a star would have been at its closest approach distance to the cluster center, but a star may have formed significantly earlier than that, and only have begun moving away from the cluster center following residual gas expulsion around the cluster and the loss of binding mass. Therefore, isochronal age estimates should be higher than kinematic age estimates for stars that originated within the cluster, and the average difference indicates the duration of an ``embedded phase'' \citep[e.g.,][]{forbidden}. In Figures ~\ref{Fig:N2264ages},~\ref{Fig:S2264ages} we can see that this holds true for most YSOs in NGC 2264 N \& S, as generally fewer stars are in the greyed out ``forbidden'' region, where isochronal age is less than kinematic age. However, many YSOs belonging to Collinder 95 are located in the 'forbidden zone', with kinematic ages seemingly too old for their isochronal ages. 

Comparing the isochronal ages for each cluster using each stellar isochrone model, in general the PARSEC \citep{parsec} models give older ages, followed by SPOTS \citep{somers20} models, and the \citet{baraffe} models give the youngest. 


We also compare ages for subsets of cluster members filtered on closest approach distance in Figs.~\ref{Fig:N2264ages},~\ref{Fig:S2264ages},~\ref{Fig:Col95ages}, requiring the closest approach distance $< r_c$ (\textit{lower rows}) or closest approach distance $< r_c$ within the 1$\sigma$ uncertainty (\textit{middle rows}), and that the YSO is located outside the core radius $r_c$. 
In Figs.~\ref{Fig:consistent_N2264ages},~\ref{Fig:consistent_S2264ages},~\ref{Fig:consistent_coll95ages}, half-mass radius ($r_{50}$) is used rather than core-radius ($r_c$), with the additional requirement that the YSOs' closest approach distance  is within $r_{50}$, while their current distance is outside of $r_{50}$. (In Fig. ~\ref{Fig:both_all_isochronal_ages} some YSOs have BP-RP $\tau_{\rm iso,  SPOTS} = 0$, meaning that many are located above the SPOTS $\log (\rm Age) =4$ isochrone—likely because their extinction is not well estimated or they are binaries. Nevertheless, these younger ages ($< 0.5$ Myr) are unreliable. There are fewer zeros in the G-RP isochronal ages than in the BP-RP isochronal ages; therefore, G-RP is used in Figs. ~\ref{Fig:N2264ages},~\ref{Fig:S2264ages}, \& ~\ref{Fig:Col95ages}.)

For both NGC 2264 N \& S clusters the majority of cluster members have isochronal ages greater than their kinematic ages. However, for Collinder 95 there are still a significant number of member YSOs with $\tau_{\rm kin,CA}>\tau_{\rm iso}$ for any stellar evolution model, which thus inhabit the ``forbidden'' zone. A possible interpretation of this would be that YSOs in the ``forbidden'' zone did not originate from their position of closest approach to the cluster center, but from an initially sparser distribution, and thus $\tau_{\rm kin,CA}$ are overestimated ages. Alternatively, there is also the possibility that all of our isochronal age estimates are underestimated, either due to underestimated reddening, or otherwise that the stellar evolution models themselves systematically underestimate ages for YSOs in this mass range.

Furthermore, instead of comparing the kinematic ages and isochronal ages of individual cluster members, we can also consider the cluster as a whole. We find that the median kinematic age values for all three clusters are lower than the median isochronal age values in nearly all cases, as seen in Figure \ref{Fig:both_all_isochronal_ages}, for both the \citet{baraffe} and PARSEC isochrones. Here, we can see clearly that the \citet{baraffe} isochrones yield much lower ages than the PARSEC isochrones.
Furthermore, in Figure \ref{Fig:both_all_isochronal_ages} we also note again that, as expected, very few cluster medians fall in the grey ``forbidden'' zone where isochronal ages are less than kinematic ages \citep[see also][]{forbidden}.

\begin{table*}[htbp]
\footnotesize
\begin{tabular}{@{} *{9}{>{\centering\arraybackslash}p{1.65cm}} @{} } 
\toprule




Star Cluster & YSO Subpopulation & Median Kinematic Age (Myr) & Median BP-RP \citet{baraffe} Isochronal Age (Myr) & Median G-RP \citet{baraffe} Isochronal Age (Myr) & Median BP-RP PARSEC Isochronal Age (Myr) & Median G-RP PARSEC Isochronal Age (Myr)  & Median BP-RP SPOTS Isochronal Age (Myr) & Median G-RP SPOTS Isochronal Age (Myr) \\  
\midrule \addlinespace
 NGC 2264 N & All YSOs & 0.77 & 2.72 & 1.42 & 4.02 & 8.74 & 0.94 & 3.17 \\
    & Within $\pm50$\textdegree & 1.08 & 2.85 & 1.49 & 3.86 & 6.72 & 1.25 & 2.85 \\
    & Travelled Significantly Outward and within $\pm50$\textdegree & 1.81 & 3.23 & 2.00 & 4.27 & 5.08 & 1.71 & 3.13 \\
\midrule \addlinespace
 NGC 2264 S & All YSOs & 0.58 & 2.72 & 1.42 & 4.02 & 8.74 & 0.94 & 3.17 \\
    & Within $\pm50$\textdegree & 0.61 & 2.95 & 1.84 & 4.32 & 6.72 & 1.73 & 3.18 \\
    & Travelled Significantly Outward and within $\pm50$\textdegree & 1.14 & 2.99 & 2.00 & 4.33 & 6.21 & 1.81 & 2.82 \\
\midrule \addlinespace
 Collinder 95 & All YSOs & 0.67 & 2.13 & 0.62 & 3.84 & 11.22 & 0.00 & 2.74 \\
    & Within $\pm50$\textdegree & 0.94 & 1.97 & 0.59 & 3.86 & 12.17 & 0.00 & 2.36 \\
    & Travelled Significantly Outward and within $\pm50$\textdegree & 1.62 & 1.59 & 0.60 & 3.86 & 10.99 & 0.00 & 2.46 \\
\bottomrule
\end{tabular}
\caption{Median kinematic age values for clusters NGC 2264 N, NGC 2264 S, and Collinder 95. Median values of isochronal ages are also included for comparison. Breakdowns across two YSO subpopulations are included as well. The first row of median values includes all YSOs when calculating the cluster median cluster age. The second represents a subpopulation and includes only YSOs with a proper motion within $\pm 50 $\textdegree$ $ of the radial. The third median includes only YSOs which both have a proper motion within $\pm 50 $\textdegree$ $ of the radial and have traveled significantly outwards from the cluster center over their lifetimes ($R_{\text{closest approach}} < \text{half-mass radius}$ and $R_{\text{current}} > \text{half-mass radius}$.) In general, the cluster median kinematic ages are significantly lower than isochronal ages.}
    \label{tab:all_vals}
\end{table*}


\subsection{Age Spread}

In each cluster, we look for evidence of significant spread in each of $\tau_{\rm kin,CA}$ and $\tau_{\rm iso}$ estimated using each set of stellar evolution models. Following the same approach as \citet{lambda_ori}, we calculate the difference between $\tau$ for individual cluster members and the sample mean $\bar{\tau}$ and normalize by the uncertainty of $\tau$ for each cluster member, removing the largest 10\% as outliers. We report the median normalised age differences, $\sigma_{\rm \tau,kin,CA}$ and $\sigma_{\rm \tau,iso}$, per cluster.

For all YSOs consistent with moving away from the cluster centers, which thus have a valid traceback age to closest approach $\tau_{\rm kin,CA}$, we obtain spreads in kinematic age $\sigma_{\rm \tau,kin,CA}$ of 23.6, 29.0 and 9.9 for NGC 2264 N, NGC 2264 S, and Collinder 95, respectively. For these stars we obtain spreads in $\tau_{\rm iso}$ using Baraffe models with BP-RP color of 22.3, 21.1 and 20.4, but only 3.4, 2.2 and 2.1 with G-RP color. With PARSEC models we obtain $\sigma_{\rm \tau,iso}$ of 12.8, 10.9 and 9.3 with BP-RP, and 2.1, 2.1 and 4.2 with G-RP. With SPOTS models we obtain $\sigma_{\rm \tau,iso}$ of 6.1, 6.9 and 8.9 with BP-RP, and 2.3, 1.9 and 2.9 with G-RP. 

For YSOs whose closest approach distance is within the cluster core radius, $r_c$, we obtain $\sigma_{\rm \tau,kin,CA}$ of 5.3, 116.7 and 46.4 for NGC 2264 N, NGC 2264 S, and Collinder 95, respectively. For these stars we obtain spreads in $\tau_{\rm iso}$ using Baraffe models with BP-RP color of 46.7, 8.4 and 24.1, and 2.5, 2.0 and 2.9 with G-RP color. With PARSEC models we obtain $\sigma_{\rm \tau,iso}$ of 23.0, 4.7 and 17.1 with BP-RP, and 2.0, 2.5 and 1.2 with G-RP. With SPOTS models we obtain $\sigma_{\rm \tau,iso}$ of 13.0, 4.3 and 23.7 with BP-RP, and 1.9, 1.8 and 4.7 with G-RP. 

For YSOs whose closest approach distance is within 1$\sigma$ uncertainty of the cluster core radius, $r_c$, we obtain $\sigma_{\rm \tau,kin,CA}$ of 6.0, 12.9 and 6.4 for NGC 2264 N, NGC 2264 S, and Collinder 95, respectively. For these stars we obtain spreads in $\tau_{\rm iso}$ using Baraffe models with BP-RP color of 24.9, 24.9 and 14.5, and 2.0, 2.2 and 2.5 with G-RP color. With PARSEC models we obtain $\sigma_{\rm \tau,iso}$ of 10.5, 6.3 and 8.5 with BP-RP, and 1.7, 3.4 and 3.9 with G-RP. With SPOTS models we obtain $\sigma_{\rm \tau,iso}$ of 8.2, 6.2 and 19.5 with BP-RP, and 1.9, 1.2 and 2.6 with G-RP. 

Overall, the evidence for isochronal age spread $\sigma_{\rm \tau,iso}$ is uncertain, i.e., the significance of the spread depends strongly on the models used to estimate ages and even on the photometric colors used for the same set of models, but is not strongly dependent on the filtering of cluster members by closest approach distance. 

We do find evidence for kinematic age spread $\sigma_{\rm \tau,kin,CA}$ in all clusters, and particularly $\sigma_{\rm \tau,kin,CA}$ is significant for cluster members with closest approach distances $<r_c$ of each cluster. This would likely indicate for NGC 2264 N \& S, which have dense and compact cores, that cluster members have gradually been ejected or become unbound from the cluster cores over their lifetimes (2-3 Myr). For Collinder 95, which is sparse and likely was not as compact at formation as NGC 2264 N \& S, the kinematic age spread is more likely a result of kinematic ages being overestimated for YSOs that did not originate from their closest approach positions, i.e., the same reason as why many cluster members inhabit the ``forbidden'' zones of Fig.~\ref{Fig:Col95ages}.

\subsection{Cluster Halos}
\label{Halos}

\begin{figure*} 
	\centering
	\includegraphics[width=0.99\linewidth]{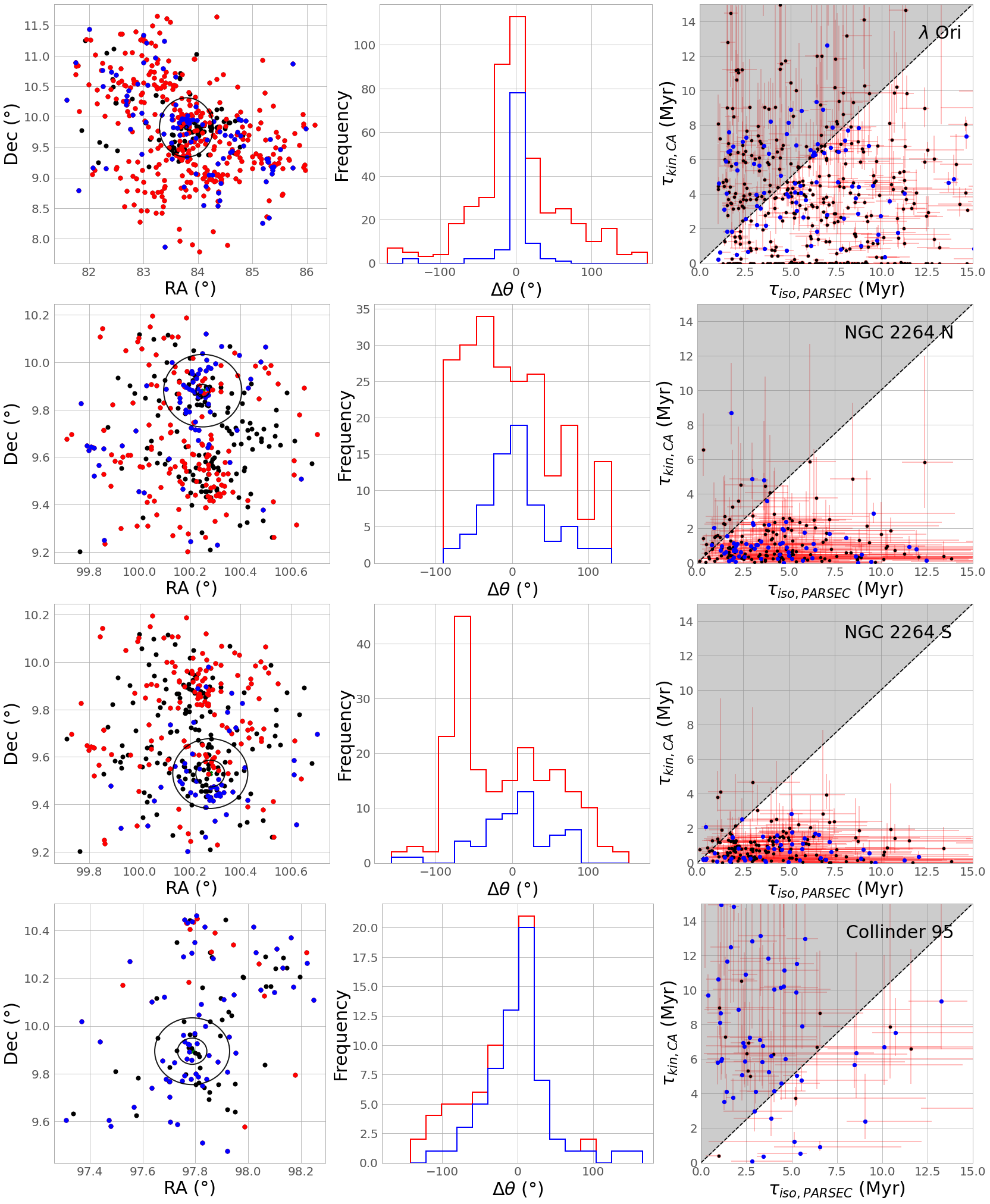}
    \caption{Spatial coordinates (\textit{left column}) with cluster core radii $r_c$ and half-mass radii $r_{50}$ indicated as black circles, relative velocity orientation $\Delta\theta$ histograms (\textit{middle column}) and PARSEC ages $\tau_{iso}$ against traceback to closest approach age $\tau_{kin,CA}$ (\textit{right column}), for $\lambda$ Ori, NGC 2264 N \& S and Collinder 95 (\textit{descending rows}). Cluster members with closest approach distance within $1\sigma$ of the core radius $r_c$ and with present position outside of the core radius are indicated as blue points.}
    \label{Fig:cluster_halos}
\end{figure*}

We compare the distributions of the clusters halos in $\lambda$ Ori \citep{lambda_ori}, NGC 2264 N \& S, and Collinder 95, defined as cluster members consistent with originating within the cluster core radii $r_c$ within $1\sigma$ of their closest approach distance (blue; Fig.~\ref{Fig:cluster_halos}). We show their spatial distributions, histograms of relative velocity orientation $\Delta\theta$ and isochronal ages, $\tau_{\rm iso}$, according to PARSEC models \citep{marigo17} using Gaia BP-RP color against traceback timescale to closest approach, $\tau_{\rm kin,CA}$ (Fig.~\ref{Fig:cluster_halos}).

The distributions of halo members for each cluster are distinct. For NGC 2264 N \& S, the majority of halo members are located within each cluster's half-mass radius $r_{50}$, have wider distributions of $\Delta\theta$, though still with peaks close to $0^\circ$, and have young traceback ages, $\tau_{\rm kin,CA}$, generally $< 3$ Myr and younger than their isochronal ages, $\tau_{\rm iso}$. For Collinder 95, the majority of halo members are located outside the cluster's half-mass radius $r_{50}$, have a narrower distribution of $\Delta\theta$ with a peak close to $0^\circ$, but have older traceback ages, $\tau_{\rm kin,CA}$, that are generally $4-14$ Myr and older than their isochronal ages, $\tau_{iso}$. For $\lambda$ Ori, close to half of halo members are located outside the cluster's half-mass radius $r_{50}$ and half are within. They have a very narrow distribution of $\Delta\theta$ with a peak close to $0^\circ$, but have a mix of traceback ages, $\tau_{\rm kin,CA}$, half of which are older than their isochronal ages, $\tau_{iso}$, and half of which are younger.

Since the vast majority of halo members of NGC 2264 N \& S have $\tau_{\rm kin,CA}<\tau_{\rm iso}$, they are consistent with having formed within the bound cores of these clusters, but have subsequently become unbound, likely including some via dynamical ejection, as the clusters evolve dynamically. Apart from these halo members, there is also a sparse distribution of YSOs surrounding the cluster cores which are not consistent with originating in either core. These belong to smaller substructures which could be considered part of the wider Mon OB1 association \citep{rapson14,wright20}.

Since the majority of halo members for Collinder 95 have $\tau_{\rm kin,CA}>\tau_{\rm iso}$, they are not consistent with having originated within the cluster core, unless the isochronal ages $\tau_{\rm iso}$ are severely underestimated. It is possible that many of these may have formed in smaller substructures surrounding the core, similar to the non-halo YSOs surrounding NGC 2264 N \& S, but then it is striking that so many of these have velocities directed radially away from the core, consistent with overall linear radial expansion and seemingly with little random motion inherited from the turbulence of their natal cloud. The large spread of $\tau_{\rm kin,CA}$ among halo members also indicates that they would have escaped the core of Collinder 95 gradually rather than after a single catastrophic event, such as a supernova expelling residual gas, or alternatively that they originated at a range of distances from the cluster core and have since moved radially away from it. 

As described in \citep{lambda_ori}, $\lambda$ Ori also has many halo members with $\tau_{\rm kin,CA}>\tau_{\rm iso}$, but most of these disappear with a more restrictive filter on closest approach distance leaving only candidate ejected stars with timescales compatible with other age estimates ($<4$ Myr), unlike with Collinder 95 (see Fig.~\ref{Fig:Col95ages}). The significant substructure surrounding the cluster core, which appears to be inconsistent with originating from it, similar to NGC 2264, has lead to the conclusion that much of this population formed sparsely distributed and hierarchically structured \citep{kounkel18,lambda_ori}.

\section{Conclusions} 

By analyzing optical stellar spectra and flagging stars with spectral indicators of youth ($|$EW(H$\alpha$)$|$ $> 10${\AA} or EW(Li) $> 1.5${\AA}), we are able to identify $166$ YSOs across several young clusters (NGC 2264 N \& S, Collinder 95, and Collinder 359). Using an additional $325$ YSOs flagged as variable stars by the \textit{Gaia} DR3 catalog \citep{gaia_var}, we generate lists of YSOs (with $N>100$) for both NGC 2264 N \& S and Collinder 95, allowing us to calculate the kinematic or traceback ages for these clusters. 
After limiting the YSOs used in the kinematic age estimation to only those with $\Delta\theta$ within $\pm50$\textdegree$ $, we find a median kinematic age of $1.08$ Myr for NGC 2264 N, $0.61$ Myr for NGC 2264 S, and $0.94$ Myr for Collinder 95.

We find in all clusters significant numbers of YSOs outside of the cluster radii with velocities inconsistent with origin within the cluster radii (see Fig.~\ref{Fig:colormap_grid}). This provides evidence for a sparse population of YSOs surrounding the dense cluster cores, with distinct substructure and kinematics, which could belong to the young stellar association Mon OB1. The existence of bound, compact clusters within the volumes of sparse associations has been observed elsewhere \citep[e.g., Vela OB2;][]{Armstrong22} 

We compare the kinematic age values with isochronal age values for the \citet{baraffe} and PARSEC \citep{parsec} stellar evolution models. 
As seen in Figure \ref{Fig:both_all_isochronal_ages}, although the three isochronal models give significantly different results, most models provide isochronal age estimates which are, on average, higher than our kinematic age estimates, which is consistent with YSOs forming within a more compact cluster volume and then subsequently becoming unbound and drifting away.

We find significant differences between ages estimated when fitting stellar evolution models to YSOs in either BP-RP or G-RP color, as can be seen in Figs.~\ref{Fig:consistent_N2264ages},~\ref{Fig:consistent_S2264ages},~\ref{Fig:consistent_coll95ages}, \&~\ref{Fig:both_all_isochronal_ages} and in ages reported in Table~\ref{tab:all_vals}. This indicates systematic differences between the observed photometric bands and the corresponding synthetic photometric bands as predicted by stellar evolution models. 

The systematic offset between kinematic ages and isochronal estimates, of 2-4 Myr according to PARSEC models in BP-RP color, for example (see Fig.~\ref{Fig:both_all_isochronal_ages}), has been suggested in similar recent works \citep[such as ][]{miret-roig24} to indicate an early period in cluster evolution where the cluster is still embedded in its parent molecular cloud for several Myrs, before stellar feedback expels this gas and cluster members begin to become unbound as the majority of binding mass is lost. However, this offset is not observed for all clusters. \citet{lambda_ori} estimated kinematic ages for the $\lambda$ Ori cluster using several approaches, but found that these were all in agreement with or even greater than ages estimated in the literature using stellar evolution models. They also found evidence that the $\lambda$ Ori cluster likely formed in a sparse, hierarchically structured configuration and so was initially unbound. 

Collinder 95, similarly, has a sparse distribution of YSO members, a majority of which have relative motions consistent with expanding away from the cluster center. Also, for those YSO members consistent with being unbound from the cluster, the median kinematic age (1.62 Myr) is in close agreement with the median isochronal age estimated using \citet{baraffe15} isochrones in BP-RP color (1.59 Myr; Table~\ref{tab:all_vals}), which would suggest that Collinder 95 formed initially sparse and unbound. However, the uncertainty in the isochronal age estimates makes this difficult to verify. This contrasts with the NGC 2264 N \& S clusters, which are more massive and densely concentrated and which likely formed bound before subsequently losing cluster members a few Myr into their evolution.

\begin{acknowledgements}
I.C. acknowledges support from a Chalmers Astrophysics \& Space Science Summer (CASSUM) Research Fellowship, including partial support from W\&M Lundgrens grant 2022-4019. J.J.A. acknowledges support from a Chalmers Initiative on Cosmic Origins (CICO) postdoctoral fellowship. J.C.T. acknowledges support from ERC Advanced Grant MSTAR (788829). This work has made use of data from the ESA space mission {\it Gaia} (http://www.cosmos.esa.int/gaia), processed by the Gaia Data Processing and Analysis Consortium (DPAC, http://www.cosmos.esa.int/web/gaia/dpac/consortium). Funding for DPAC has been provided by national institutions, in particular the institutions participating in the Gaia Multilateral Agreement. This research made use of the Simbad and Vizier catalogue access tools (provided by CDS, Strasbourg, France), Astropy \citep{astr13} and TOPCAT \citep{tayl05}.
\end{acknowledgements}

\bibliography{aanda}{}
\bibliographystyle{aa}

\begin{appendix} 
\section{Supplementary Figures}

\begin{figure*} 
	\centering
	\includegraphics[width=\linewidth]{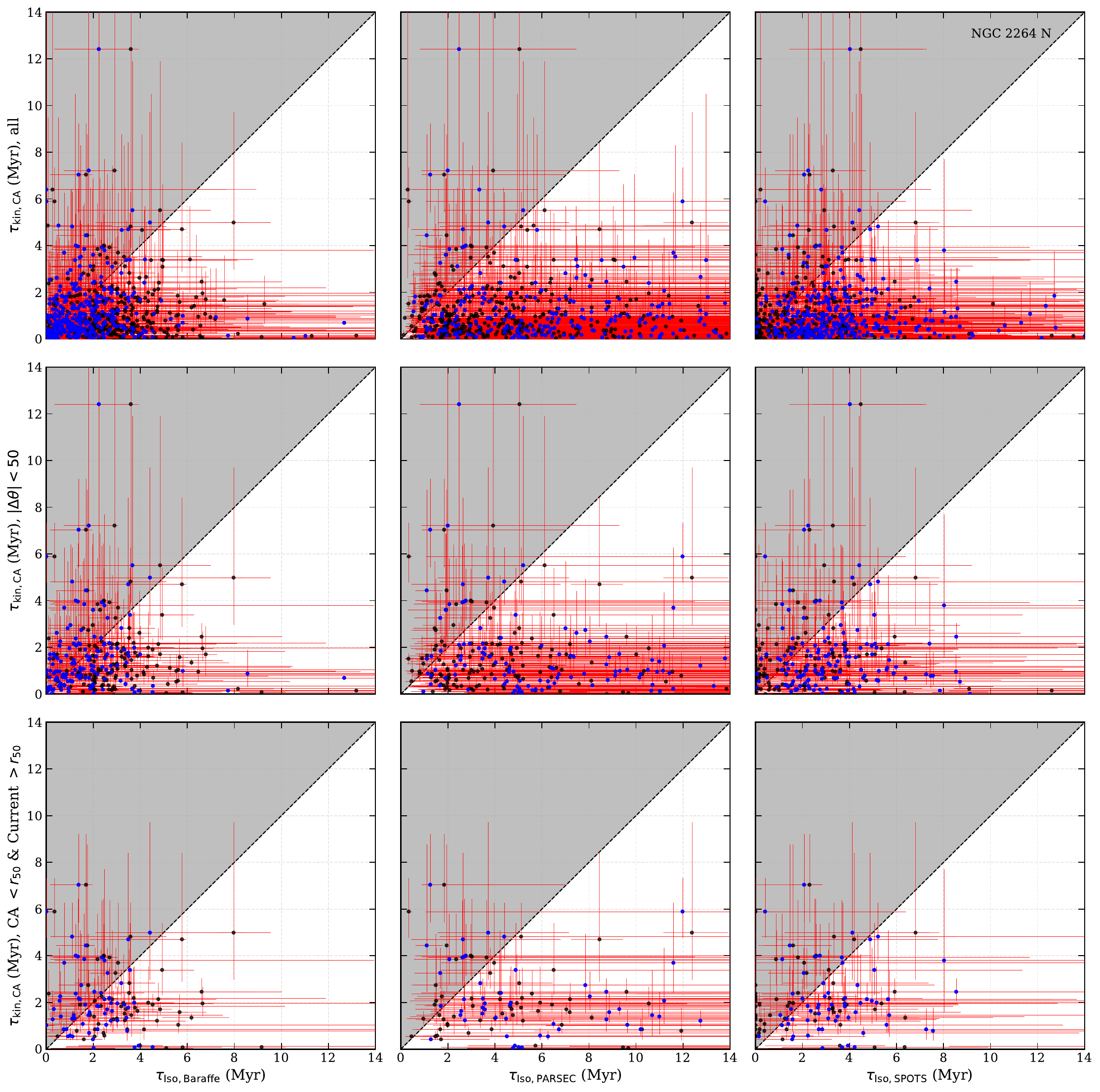}
    \caption{Kinematic age, $\tau_{\rm kin}$, against isochronal age, $\tau_{\rm iso}$, for all YSOs (\textit{top panel}), for YSOs moving away from the cluster center (\textit{middle panel}), and for YSOs with closest approach distances within $r_{50}$ and current distances outside of $r_{50}$ (\textit{lower panel}), for NGC 2264 N. The plot denotes photometric color used for $\tau_{\rm iso}$ using point color (BP-RP as black, G-RP as blue.)}
    \label{Fig:consistent_N2264ages}
\end{figure*}

\begin{figure*} 
	\centering
	\includegraphics[width=\linewidth]{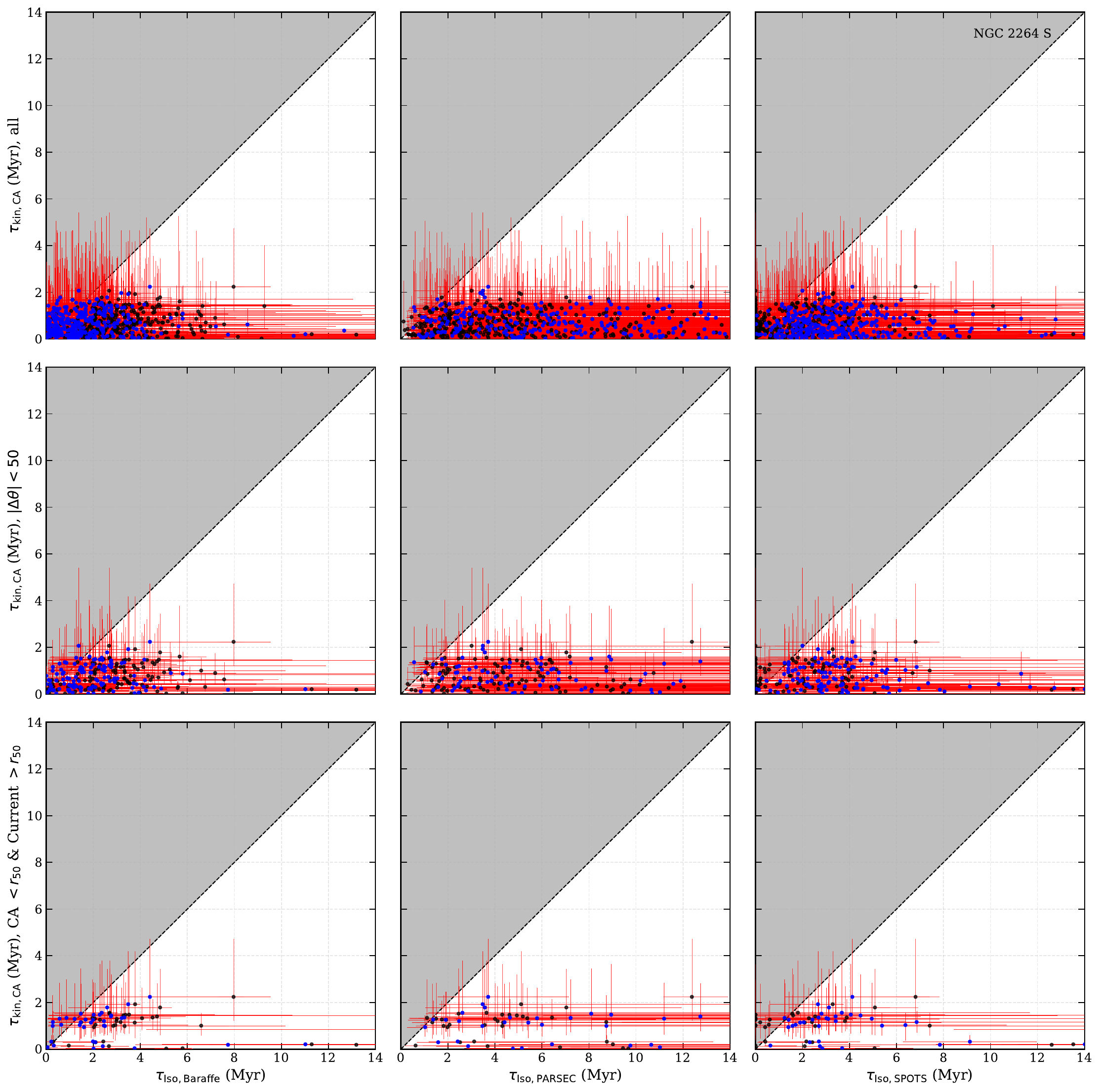}
    \caption{As Fig.~\ref{Fig:consistent_N2264ages}, but for NGC 2264 S.}
    \label{Fig:consistent_S2264ages}
\end{figure*}

\begin{figure*} 
	\centering
	\includegraphics[width=\linewidth]{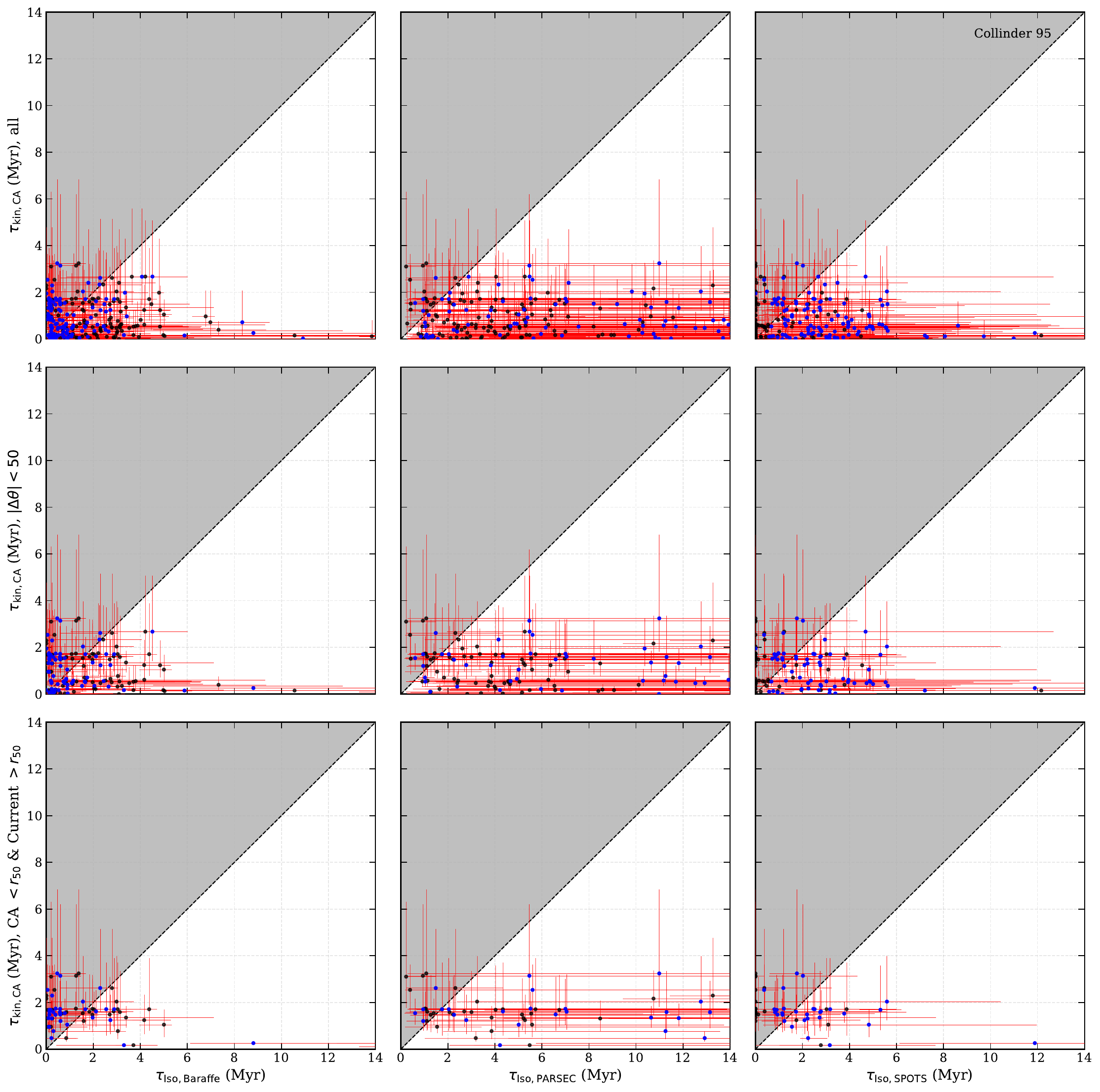}
    \caption{As Fig.~\ref{Fig:consistent_N2264ages}, but for Collinder 95.}
    \label{Fig:consistent_coll95ages}
\end{figure*}

\begin{figure*} 
	\centering
	\includegraphics[width=\linewidth]{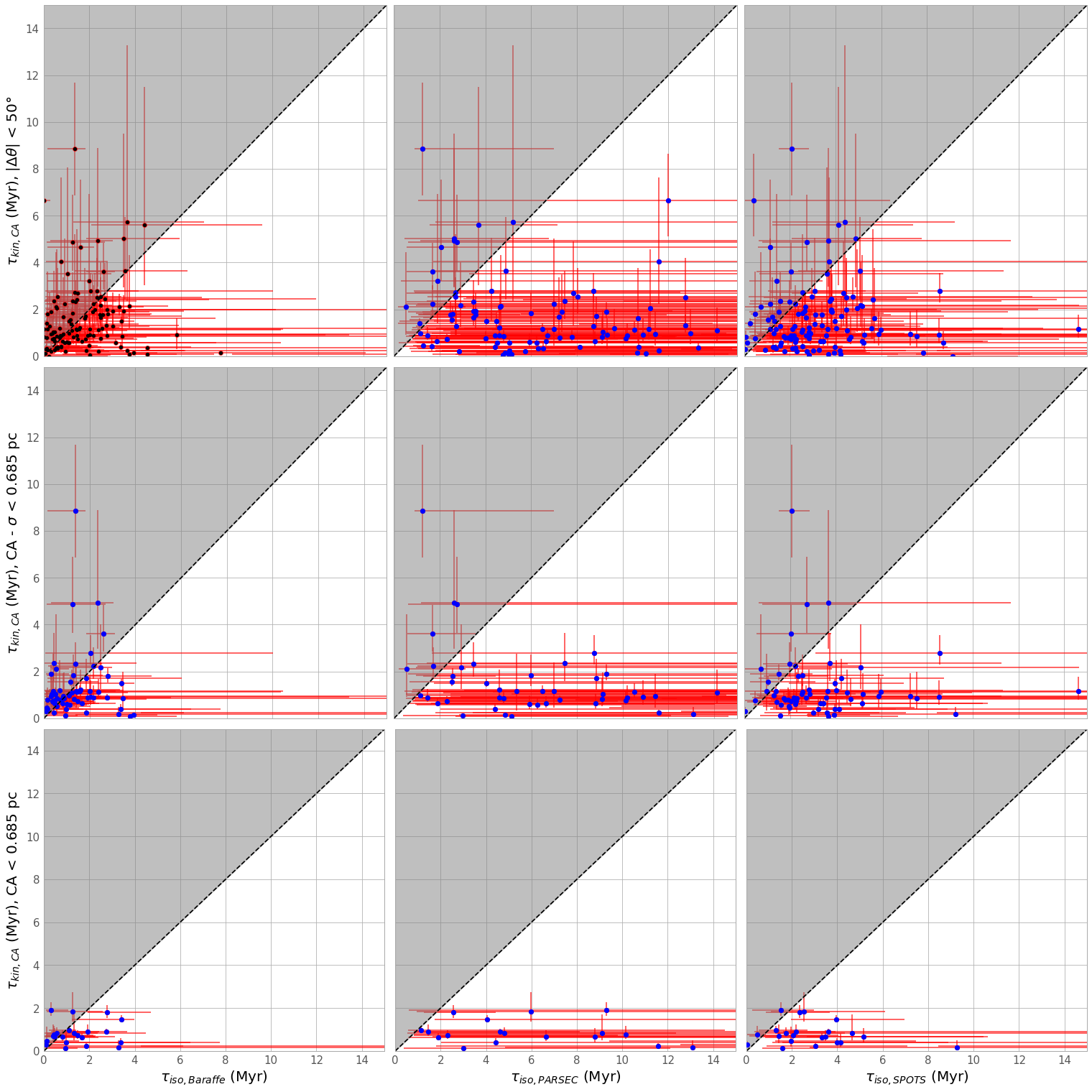}
    \caption{Kinematic age, $\tau_{\rm kin}$, against isochronal age, $\tau_{\rm iso}$ (G-RP), for YSOs moving away from the cluster center (\textit{top panel}), for YSOs with closest approach distances within 1$\sigma$ of the core radius, $r_c$ (\textit{middle panel}), and for YSOs with closest approach distances within the core radius $r_c$ (\textit{lower panel}), for NGC 2264 N.}
    \label{Fig:N2264ages}
\end{figure*}

\begin{figure*} 
	\centering
	\includegraphics[width=\linewidth]{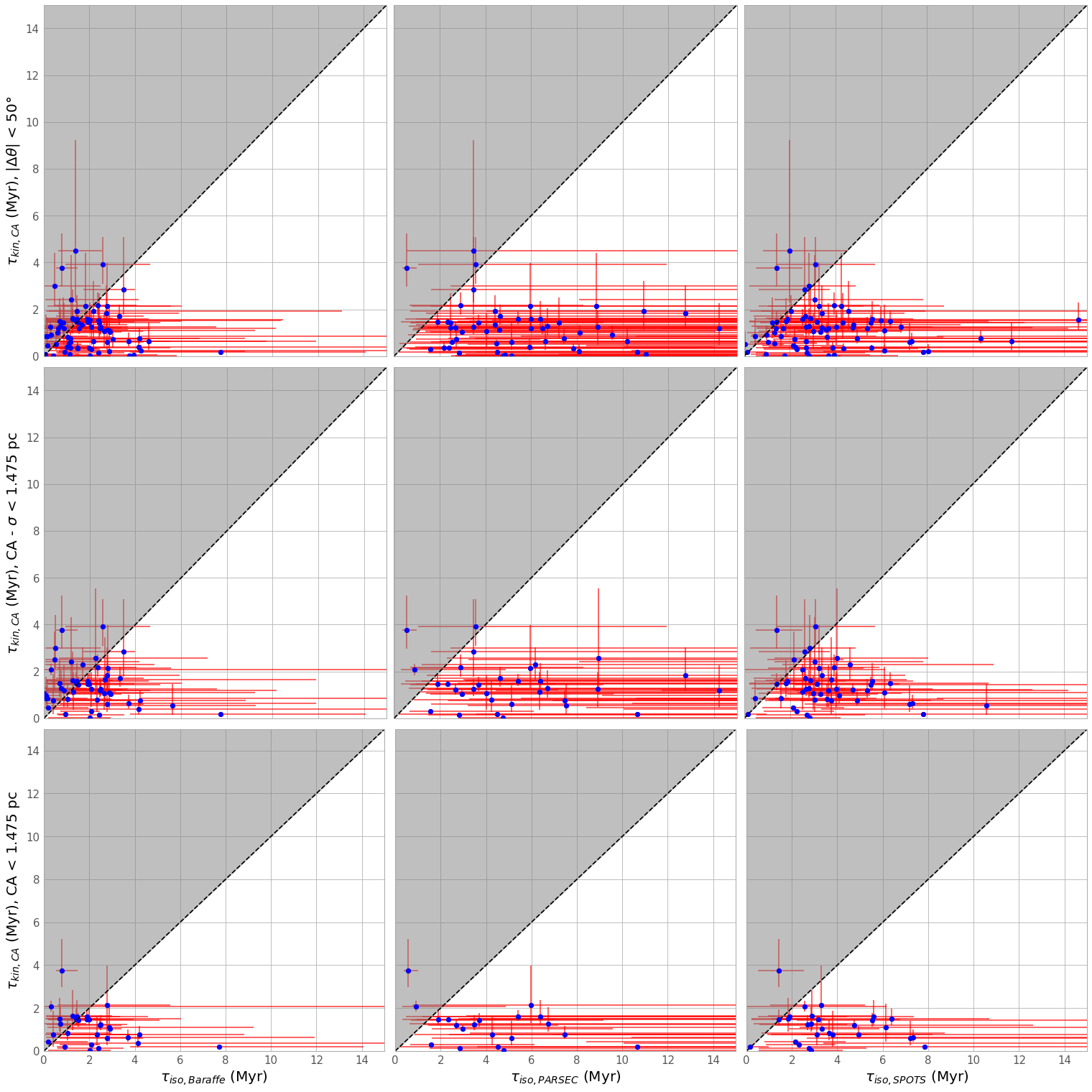}
    \caption{As Fig.~\ref{Fig:N2264ages}, but for NGC 2264 S.
    }
    \label{Fig:S2264ages}
\end{figure*}

\begin{figure*} 
	\centering
	\includegraphics[width=\linewidth]{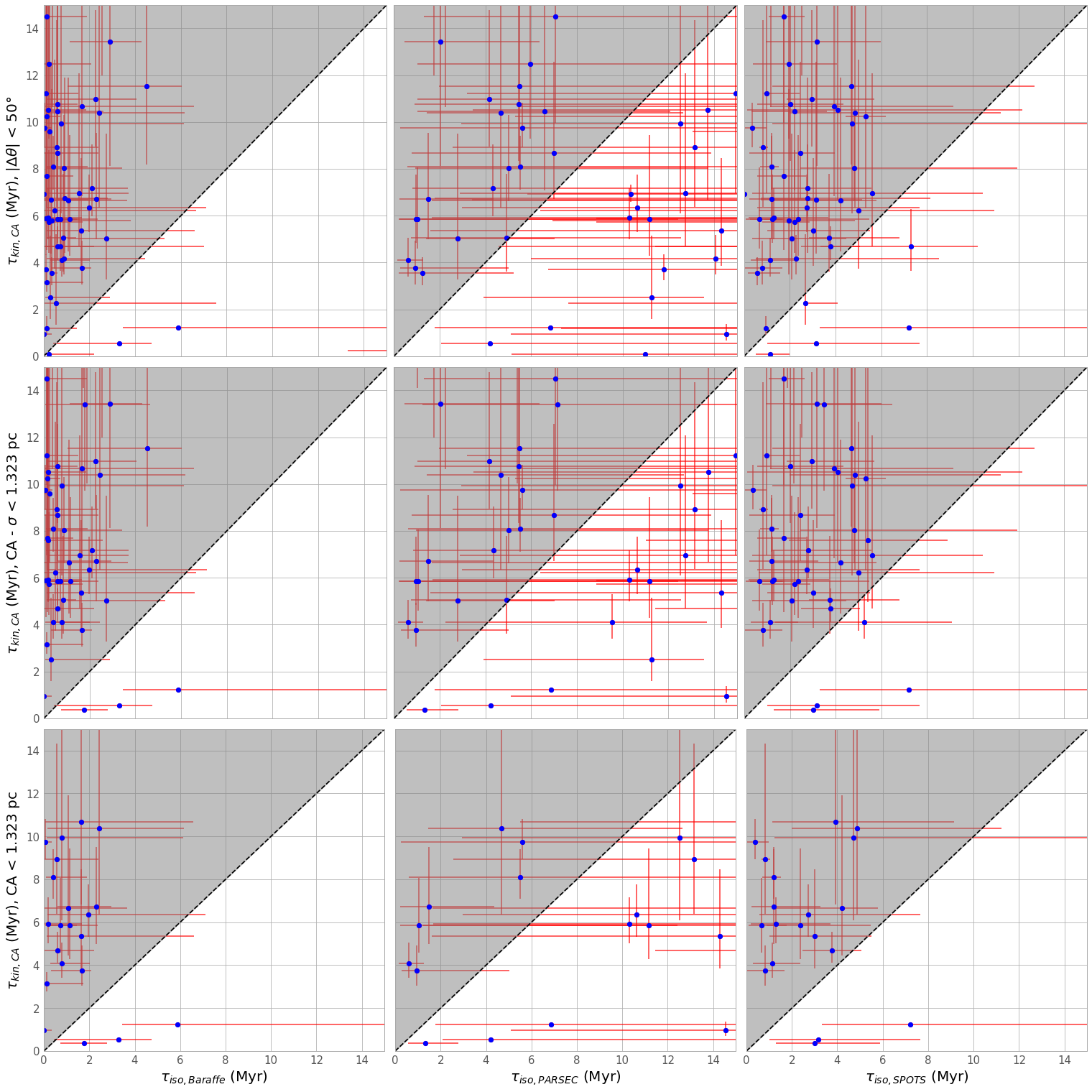}
    \caption{As Fig.~\ref{Fig:N2264ages}, but for Collinder 95.
    }
    \label{Fig:Col95ages}
\end{figure*}

\begin{figure*} 
	\centering
	\includegraphics[width=\linewidth]{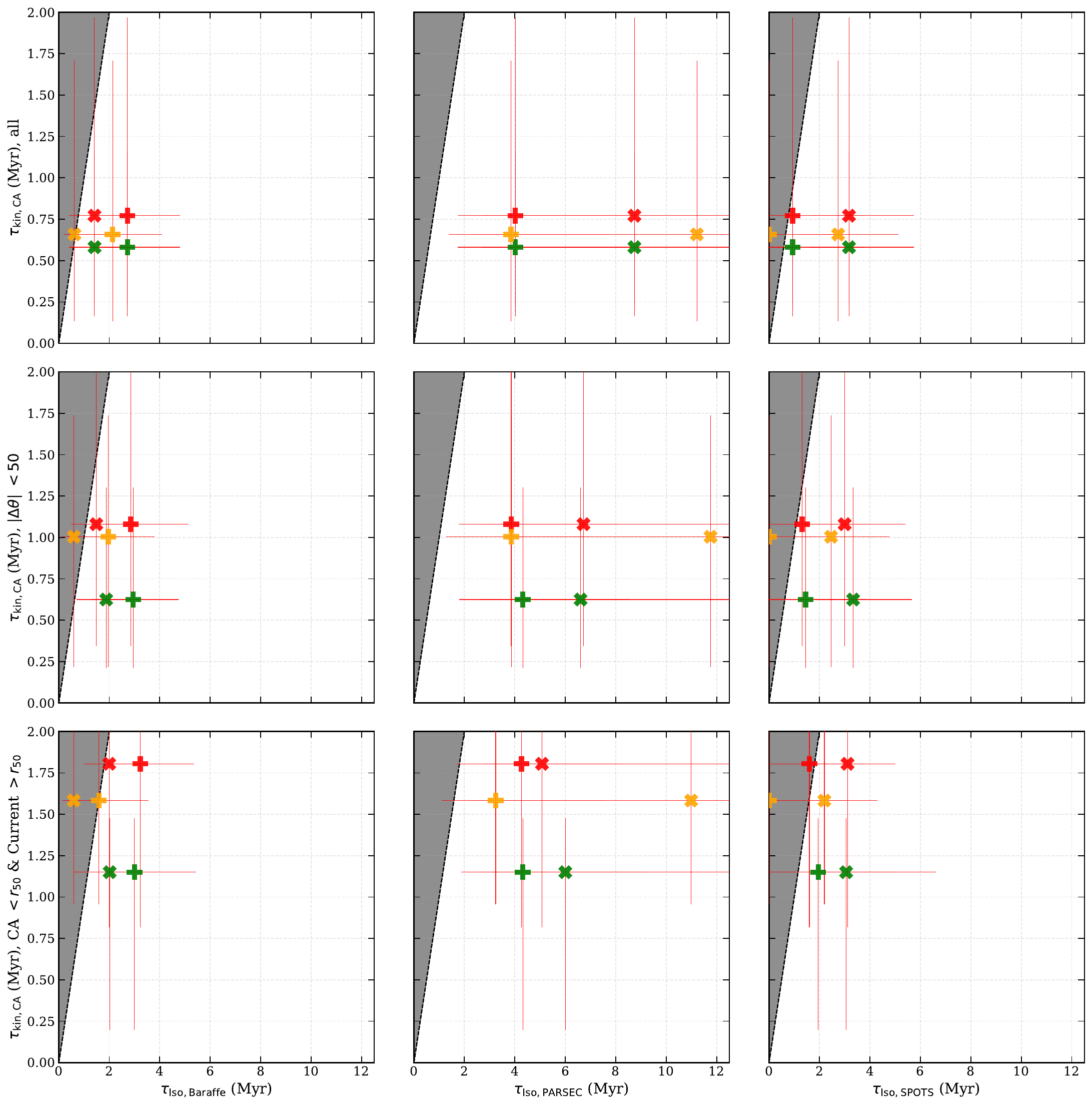}
    \caption{The median kinematic ages ($\tau_{\textrm{kin}}$) versus their median isochronal ages ($\tau_{\textrm{iso}}$) for clusters NGC2264 N (red), NGC 2264 S (green), and Coll 95 (orange). The plot is divided into nine panels, with the three subgroups as rows: all YSOs (\textit{top panel}), YSOs moving away from the cluster center (\textit{middle panel}), and YSOs with closest approach distances within $r_{50}$ and current distances outside of $r_{50}$ (\textit{lower panel}). Columns denote the different isochrone models used, with Baraffe (\textit{left}), PARSEC (\textit{middle}), and SPOTS (\textit{right}). The plot denotes photometric color used (BP-RP or G-RP) using symbols (BP-RP marked by a plus, G-RP marked by an X.) The error bars on each value are the 16th and 84th percentiles of the distribution of either kinematic or isochronal ages.}

    \label{Fig:both_all_isochronal_ages}
\end{figure*}

\end{appendix}

\end{document}